\documentclass{article}

\usepackage{amsmath,amssymb,amsfonts}
\usepackage{algorithmic}
\usepackage{graphicx}
\usepackage{textcomp}
\usepackage{caption,subcaption}
\usepackage{multirow}

\usepackage[utf8]{inputenc}
\usepackage[english]{babel}
\usepackage{csquotes}
\usepackage[overload]{textcase} 

\usepackage{bm}

\usepackage[symbol]{footmisc}

\usepackage{hyperref}

\renewcommand{\thefootnote}{\fnsymbol{footnote}}

\usepackage[style = ieee]{biblatex}
\newcommand\blfootnote[1]{%
  \begingroup
  \renewcommand\thefootnote{}\footnote{#1}%
  \addtocounter{footnote}{-1}%
  \endgroup
}

\usepackage{arxiv}

\usepackage[utf8]{inputenc} 
\usepackage[T1]{fontenc}    
\usepackage{hyperref}       
\usepackage{url}            
\usepackage{booktabs}       
\usepackage{amsfonts}       
\usepackage{nicefrac}       
\usepackage{microtype}      
\usepackage{lipsum}		
\usepackage{graphicx}
\usepackage{doi}

\usepackage{tikz}
\usepackage{textcomp}
\newcommand\copyrighttext{%
\footnotesize \textcopyright 2022 IEEE. Personal use of this material is permitted.
Permission from IEEE must be obtained for all other uses, in any current or future
media, including reprinting/republishing this material for advertising or promotional
purposes, creating new collective works, for resale or redistribution to servers or
lists, or reuse of any copyrighted component of this work in other works.

\doi{10.1109/ACCESS.2022.3145370}}
\newcommand\copyrightnotice{%
\begin{tikzpicture}[remember picture,overlay]
\node[anchor=south,yshift=20pt] at (current page.south) {\fbox{\parbox{\dimexpr\textwidth-\fboxsep-\fboxrule\relax}{\copyrighttext}}};
\end{tikzpicture}%
}

\title{Efficient 3-D Near-Field MIMO-SAR Imaging for Irregular Scanning Geometries}

\author{ \href{https://orcid.org/0000-0002-3388-4805}{\includegraphics[scale=0.06]{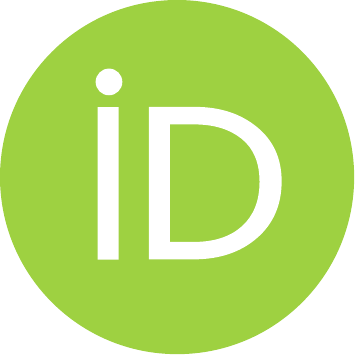}\hspace{1mm}Josiah W. Smith} \\
	Department of Electrical and Computer Engineering\\
	The University of Texas at Dallas\\
	Richardson, TX 75080 \\
	\texttt{josiah.smith@utdallas.edu} \\
	\And
	\href{https://orcid.org/0000-0001-7229-1765}{\includegraphics[scale=0.06]{orcid.pdf}\hspace{1mm}Murat Torlak} \\
	Department of Electrical and Computer Engineering\\
	The University of Texas at Dallas\\
	Richardson, TX 75080 \\
	\texttt{torlak@utdallas.edu} \\
}

\date{}

\renewcommand{\headeright}{\textit{IEEE Access} (Accepted)}
\renewcommand{\undertitle}{\textit{IEEE Access} (Accepted)}
\renewcommand{\shorttitle}{Efficient 3-D Near-Field MIMO-SAR}

\hypersetup{
pdftitle={Efficient 3-D Near-Field MIMO-SAR Imaging for Irregular Scanning Geometries},
pdfsubject={eess.SP, cs.CV},
pdfauthor={Josiah W.~Smith, Murat Torlak},
pdfkeywords={5G, automotive SAR, drone mmWave imaging, freehand imaging, handheld scanner, irregular sampling, mmWave imaging, multistatic imaging, real-time imaging, synthetic aperture radar (SAR)},
}

\begin{document}
\maketitle
\copyrightnotice

\begin{abstract}
In this article, we introduce a novel algorithm for efficient near-field synthetic aperture radar (SAR) imaging for irregular scanning geometries. 
With the emergence of fifth-generation (5G) millimeter-wave (mmWave) devices, near-field SAR imaging is no longer confined to laboratory environments. 
Recent advances in positioning technology have attracted significant interest for a diverse set of new applications in mmWave imaging. 
However, many use cases, such as automotive-mounted SAR imaging, unmanned aerial vehicle (UAV) imaging, and freehand imaging with smartphones, are constrained to irregular scanning geometries. 
Whereas traditional near-field SAR imaging systems and quick personnel security (QPS) scanners employ highly precise motion controllers to create ideal synthetic arrays, emerging applications, mentioned previously, inherently cannot achieve such ideal positioning. 
In addition, many Internet of Things (IoT) and 5G applications impose strict size and computational complexity limitations that must be considered for edge mmWave imaging technology. 
In this study, we propose a novel algorithm to leverage the advantages of non-cooperative SAR scanning patterns, small form-factor multiple-input multiple-output (MIMO) radars, and efficient monostatic planar image reconstruction algorithms. 
We propose a framework to mathematically decompose arbitrary and irregular sampling geometries and a joint solution to mitigate multistatic array imaging artifacts. 
The proposed algorithm is validated through simulations and an empirical study of arbitrary scanning scenarios. 
Our algorithm achieves high-resolution and high-efficiency near-field MIMO-SAR imaging, and is an elegant solution to computationally constrained irregularly sampled imaging problems. 
\end{abstract}

\blfootnote{This work was supported by the Semiconductor Research Corporation (SRC) task 2712.029 through The University of Texas at Dallas' Texas Analog Center of Excellence (TxACE).}


\section{Introduction}
\label{sec:introduction}
Low-cost electromagnetic imaging systems have gained attention over the past decade as commercially available radar platforms have become increasingly affordable. 
Millimeter-wave (mmWave) radar has attracted exceptional interest for applications including gesture recognition \cite{smith2021sterile}, concealed threat detection \cite{sheen2016three,yanik2019near}, and medical imaging \cite{chao2012millimeter,gao2016millimeter} due to its semi-penetrating non-ionizing nature and low power consumption. 
Additionally, with the emergence of fifth-generation (5G) and sixth-generation (6G) technology, ultra-wideband (UWB) mmWave transceivers are enabling unprecedented sensing and communications feats \cite{li2021Integrated,alvarez2021towards,basrawi2021reverse}.
Small form-factor multiple-input-multiple-output (MIMO) radars are increasing in popularity due to low cost and power consumption \cite{smith2021An,yanik2019near}. 
In addition to emerging 5G communications, mmWave radar has already been realized for high-resolution sensing on the Google Pixel 4 \cite{basrawi2021reverse}.
Of particular interest, recent work has enabled freehand mmWave imaging by employing positioning sensors commonly employed in smartphones and virtual reality (VR) sensor suites \cite{alvarez2021towards,alvarez2019freehand,alvarez2021freehand,alvarez2021freehandsystem,alvarez2021system}.
Sub-wavelength localization accuracy was previously unachievable by conventional techniques such as 5G mmWave \cite{wymeersch2017mmWavePositioning} or Bluetooth low energy (BLE) ranging \cite{hajiakhondi2020bluetooth}.
Freehand mmWave imaging is a high-resolution imaging technique that relies on conventional synthetic aperture radar (SAR) principles \cite{smith2020nearfieldisar,yanik2019sparse,yanik2019cascaded,lopez20003,yanik2020development} and precise tracking of the handheld radar device as it is moved by a human user throughout space \cite{alvarez2021towards,baumgartner2017sononet,blackall2005alignment,gilbertson2015force}.
Whereas traditional mmWave SAR imaging requires precise motion systems to achieve near-ideal synthetic arrays \cite{yanik2020development}, the scanning geometry employed by freehand imaging systems is generally irregular and does not conform to the typical array geometries required for efficient image reconstruction algorithms \cite{smith2022ThzToolbox}.

While recent work has proposed a fast imaging algorithm for irregular SAR geometries using array linearization \cite{zeng2021aperturelinearization}, the proposed technique adopts a simplistic model of the array displacement and does not explore near-field multistatic effects, both of which are addressed in this study. 
However, efficient algorithms for near-field MIMO-SAR operation under irregular scanning geometries have not been explored in the literature. 

Extensive work on freehand mmWave imaging has been conducted by Laviada \textit{et al.} at the University of Oviedo \cite{alvarez2019freehand,alvarez2021freehand,alvarez2021freehandsystem,alvarez2021system,alvarez2021towards,garcia20203DSARProcessing,wu2020multilayered}.
High-precision localization systems that enable freehand SAR imaging have been investigated using an infrared camera network to accurately track the device location across time and recover electromagnetic (EM) images \cite{alvarez2019freehand}.
Their work was extended to employ an inertial measurement unit (IMU) and depth camera sensors to achieve standalone freehand imaging with promising results \cite{alvarez2021towards,alvarez2021system}.
In each of these efforts, the subject attempted to move the hand in a raster pattern to synthesize an approximately rectangular planar aperture with a linear frequency-modulated (LFM) handheld radar \cite{alvarez2019freehand,alvarez2021freehand,alvarez2021towards}. 
Due to the subject's inability to move their hand in an ideal planar trajectory and the sensitivity of the mmWave signal to sub-millimeter perturbations, the image was reconstructed using the generalized back-projection algorithm (BPA). 

Similar irregular and non-cooperative scanning geometries have been observed in unmanned aerial vehicle (UAV) SAR imaging \cite{garcia20203DSARProcessing}, nonuniform NDT \cite{wu2020multilayered}, and automotive SAR imaging \cite{kan2020automotiveSAR}. 
However, for many edge and mobile applications, limitations on power consumption and computational complexity cannot be overcome using existing approaches for irregularly sampled SAR.
Although image reconstruction algorithms have been thoroughly investigated in the literature for cooperative synthetic array geometries \cite{sheen2001three,yanik2019sparse,yanik2019near,yanik2019cascaded,yanik2020development,smith2020nearfieldisar,smith2021An,smith2021sterile,gao2018_1D_MIMO,fan2020linearMIMOArbitraryTopologies,gao2016efficient,lopez20003,amineh2019real,smith2022ThzToolbox}, widely applicable efficient near-field imaging algorithms for applications such as freehand smartphone imaging, UAV imaging, and automotive SAR imaging have not been thoroughly addressed in the existing literature. 
Furthermore, while MIMO arrays, commonly employed in commercially available radar devices, offer spatially efficient small array sizes, the MIMO-SAR operation introduces a handful of complications to the image reconstruction process and proper handling of the multistatic array is necessary to avoid imaging artifacts \cite{yanik2019sparse}.
While progress has been made towards projecting MIMO-SAR radar data to virtual single-input-single-output (SISO) monostatic data \cite{yanik2019sparse,smith2022ThzToolbox}, the analysis is performed on a coplanar assumption that does not generally hold for irregular scanning geometries.

In this article, we propose a novel image reconstruction technique for efficient near-field imaging with irregular scanning geometries, such as those present in freehand imaging, UAV SAR, or automotive scenarios.
We examine the system and signal models for UWB MIMO-SAR and develop a multi-planar multistatic approach to mathematically decompose the irregularly sampled synthetic array such that an equivalent virtual planar monostatic array can be constructed. 
This technique is the first to extend the range migration algorithm (RMA) such that non-cooperative SAR scanning and multistatic effects are simultaneously mitigated. 
The analysis in the subsequent sections provides a novel framework for decomposing irregular SAR scenarios and efficiently projecting irregular MIMO-SAR samples to a virtual planar monostatic equivalent. 
The proposed algorithm is validated through simulation and experimentation, demonstrating its robustness to arbitrary scanning patterns and low computational complexity.
A thorough study of the relationship between array irregularity and image resolution of the proposed algorithm. 
The proposed technique demonstrates high-fidelity focusing comparable to the traditional planar RMA, even under array perturbation on the order of 10s of wavelengths. 
Our solution enables the development of emerging technologies that require non-ideal SAR scanning geometries, MIMO multistatic radar, and efficient image reconstruction.

\begin{figure}[h]
    \centering
    \includegraphics[width=0.55\textwidth]{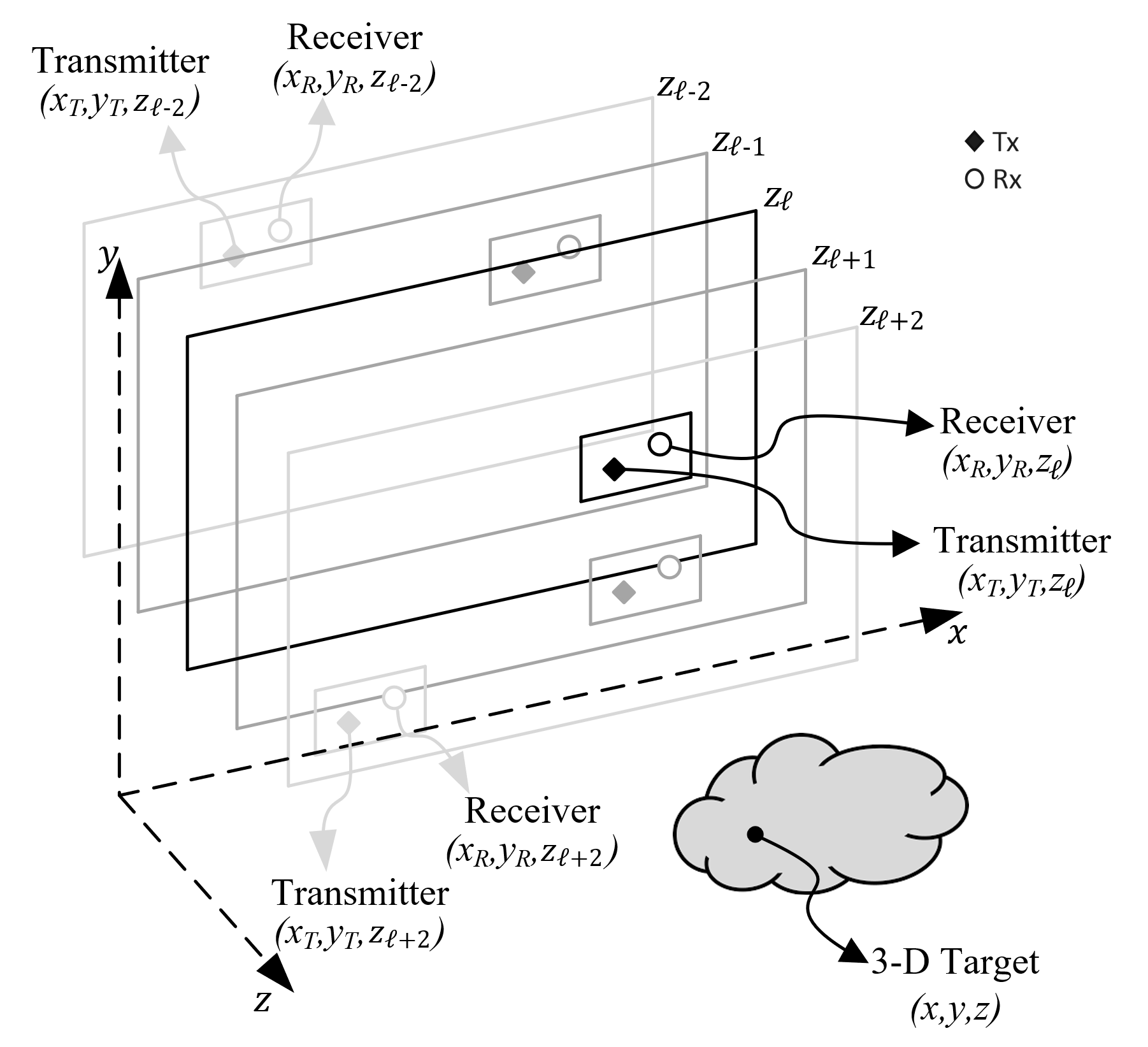}
    \caption{Geometry of the multi-planar SAR irregular scanning geometry with a multistatic array.}
    \label{fig:multiplanar}
\end{figure}

The remainder of this paper is organized as follows.
Section \ref{sec:system_model} introduces the system model, including the multi-planar multistatic SAR concept, the signal model, and a novel compensation technique to planar monostatic SAR. 
In Section \ref{sec:image_reconstruction}, efficient imaging methods and implementation details are discussed and an enhanced algorithm is proposed.
Section \ref{sec:prototype} details the hardware and software implementation for collecting multi-planar multistatic SAR data.
The results of the simulation and empirical studies are presented and discussed in Section \ref{sec:results}, followed by the conclusions.

\section{System Model}
\label{sec:system_model}
In this article, we propose the characterization of irregular or arbitrary three-dimensional (\mbox{3-D}) MIMO-SAR sampling geometry using the multi-planar multistatic scenario shown in Fig. \ref{fig:multiplanar}, where data are collected along different $z$-planes by a MIMO multistatic radar with respect to a stationary \mbox{3-D} target.

\subsection{Multi-Planar MIMO-SAR Configuration}
\label{subsec:multiplanar_SAR}
For many SAR applications mentioned in Section \ref{sec:introduction}, as the radar is moved throughout \mbox{3-D} space, it is generally oriented in the same direction towards some target; however, the samples are taken across several $z$-planes. 
As the data are collected during an arbitrary SAR scanning path, the resulting synthetic aperture does not conform to standard scanning regimes, such as rectilinear/planar \cite{sheen2016three,yanik2020development}, circular \cite{wu2020multilayered,gao2016efficient}, or cylindrical \cite{amineh2019real,smith2020nearfieldisar,smith2022ThzToolbox}.
Hence, the image reconstruction process must consider the irregularity of the spatial sampling, the geometry of which is detailed in Fig. \ref{fig:multiplanar}.

Compared with planar MIMO-SAR, which requires a multistatic MIMO array to be scanned across a planar track \cite{yanik2020development,yanik2019sparse,fan2020linearMIMOArbitraryTopologies}, multi-planar MIMO-SAR allows the multistatic array to be scanned across \mbox{3-D} space. 
For freehand imaging or automotive SAR, a MIMO array is fixed to a smartphone or vehicle, respectively, and is moved throughout space, generating a multi-planar MIMO-SAR irregular aperture. 
As shown in Fig. \ref{fig:multiplanar}, because the multistatic array is scanned in an irregular pattern spanning multiple $z$-planes, the locations of the transmit (Tx) and receive (Rx) elements are spatially translated by the movement of the MIMO array. 
The analyses in the subsequent sections present an efficient solution to irregular MIMO-SAR imaging, such that the position of the radar is known throughout the scan and the planar array assumption does not hold. 
This scenario is common for many of the aforementioned applications and necessitates both irregular scanning geometries and efficient image recovery.

\subsection{The \mbox{3-D} Multi-Planar Virtual Array Response in Near-Field Imaging}
\label{subsec:virtual_array}
\begin{figure}[h]
    \centering
    \includegraphics[width=0.55\textwidth]{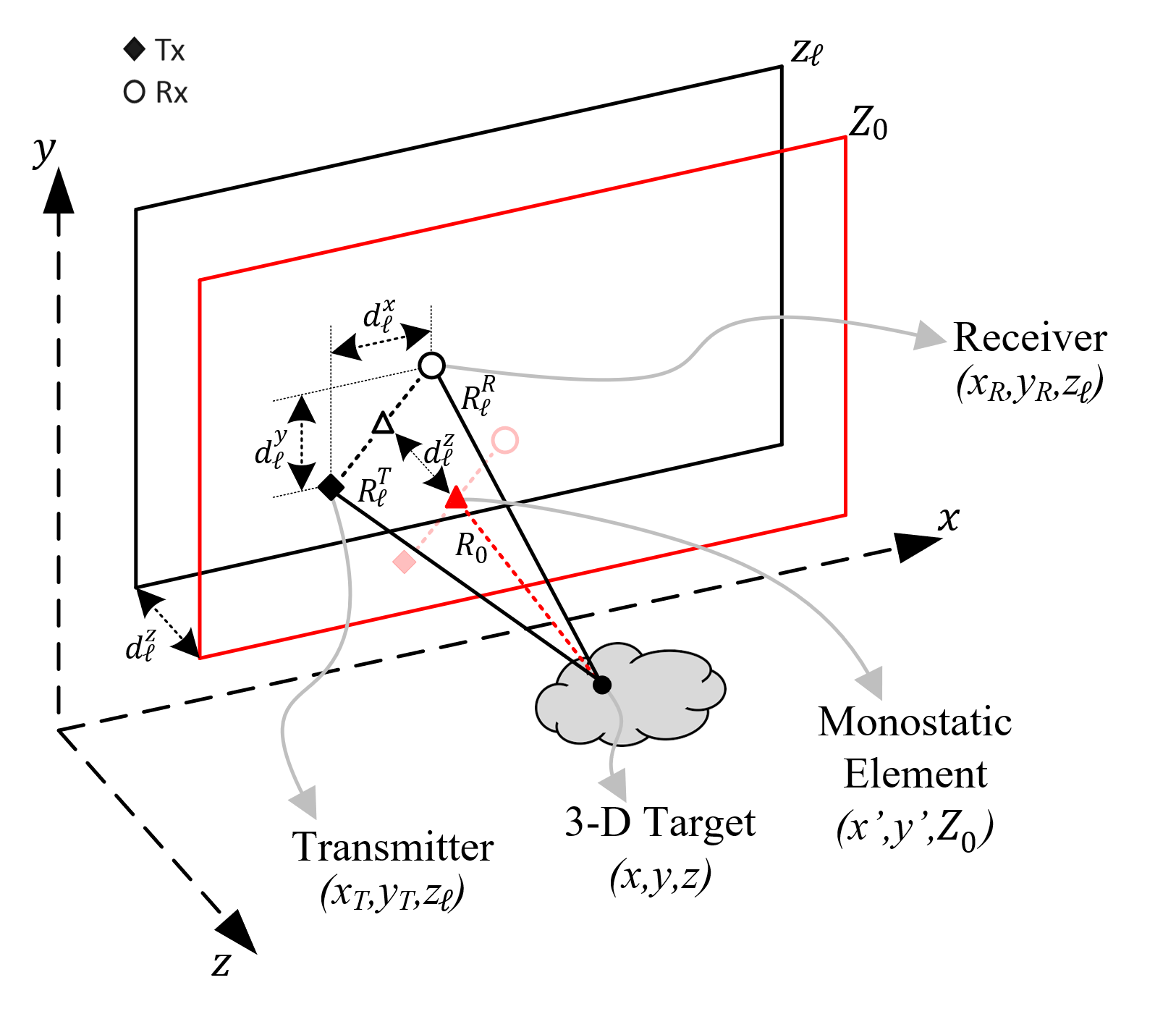}
    \caption{Relationship between the multi-planar multistatic elements and virtual planar monostatic elements.}
    \label{fig:virtual_array_compensation}
\end{figure}

By the analysis of \cite{yanik2019sparse,smith2022ThzToolbox,ender2009systemMIMOSAR} for the \mbox{2-D} case, a multistatic MIMO array can be approximated by a monostatic virtual element located at the midpoint of the Tx and Rx elements under the far-field assumption for a small fraction $\epsilon$ as 
\begin{equation}
\label{eq:far_field_assumption}
    \sqrt{(d_\ell^x)^2 + (d_\ell^y)^2} \leq \sqrt{4 \epsilon \lambda R},
\end{equation}
where $d_\ell^x$, $d_\ell^y$ are the distances between the Tx and Rx elements along the $x$- and $y$-directions, respectively, as shown in Fig. \ref{fig:virtual_array_compensation}, $\lambda$ is the wavelength of the carrier frequency, and $R$ is the distance from the midpoint of the antenna elements to a reference point in the scene. 

However, under the multi-planar multistatic framework, it is desirable to approximate each Tx/Rx pair using its virtual element located on a $Z_0$ plane in the near-field.
Thus, multi-planar data can be projected onto a virtual planar array to ease the subsequent image reconstruction process.
As shown in Fig. \ref{fig:virtual_array_compensation}, the $\ell$-th Tx/Rx pair located on the $z_\ell$ plane can be approximated by the element located at the midpoint between the Tx and Rx elements migrated to the $Z_0$ plane.

For near-field SAR, the assumption in (\ref{eq:far_field_assumption}) is invalid, and the approximation must be handled more delicately. 
Hence, we derive an efficient compensation algorithm to approximate the multistatic multi-planar array as a monostatic planar array for near-field imaging scenarios.

The transmitter (Tx) and receiver (Rx) of the $\ell$-th multistatic MIMO array are located at $(x_T,y_T,z_\ell)$ and $(x_R,y_R,z_\ell)$, respectively, and the target scene is assumed to be a distributed target whose coordinates are given by $(x,y,z)$. 
In this study, orthogonality is leveraged across time by operating a MIMO radar using the time-division multiplexing (TDM) MIMO technique such that each Tx/Rx pair is activated sequentially. 
The round-trip distance between the $\ell$-th Tx/Rx pair and the point scatter located at $(x,y,z)$ can be written as 
\begin{align}
\label{eq:Rl_of_xT_xR_yT_yR}
    \begin{split}
        R_\ell^{RT}&= R_\ell^T + R_\ell^R, \\
        R_\ell^T &= \left[(x_T - x)^2 + (y_T - y)^2 + (z_\ell - z)^2 \right]^{\frac{1}{2}}, \\
        R_\ell^R &= \left[(x_R - x)^2 + (y_R - y)^2 + (z_\ell - z)^2 \right]^{\frac{1}{2}}.
    \end{split}
\end{align}

Denoting the virtual antenna element locations as $(x',y',Z_0)$, the $x$- and $y$-coordinates of the Tx/Rx pair can be expressed as
\begin{align}
\label{eq:xT_xR_yT_yR_to_virtual}
    \begin{split}
        x_T &= x' - d_\ell^x/2, \quad y_T = y' - d_\ell^y/2, \\
        x_R &= x' + d_\ell^x/2, \quad y_R = y' + d_\ell^y/2.
    \end{split}
\end{align}

Similarly, denoting $d_\ell^z$ as the distance between the $Z_0$ plane and the $z_\ell$ plane, as shown in Fig. \ref{fig:virtual_array_compensation}, the $z$-coordinate of the Tx and Rx elements can be expressed with respect to $Z_0$ as
\begin{equation}
\label{eq:zl_to_virtual}
    z_\ell = Z_0 + d_\ell^z.
\end{equation}

As described in Appendix \ref{app:taylor_series}, substituting (\ref{eq:xT_xR_yT_yR_to_virtual}) and (\ref{eq:zl_to_virtual}) into (\ref{eq:Rl_of_xT_xR_yT_yR}) and applying the third-order Taylor series expansion of $R_\ell$ for small values of $d_\ell^x$, $d_\ell^y$, and $d_\ell^z$ yields
\begin{multline}
\label{Rl_approximation1}
    R_\ell^{RT}\approx 2R_0 + \frac{2(Z_0-z)d_\ell^z}{R_0} + \frac{(d_\ell^x)^2 + (d_\ell^y)^2 + 4(d_\ell^z)^2}{4R_0} - \frac{\left[(x'-x)d_\ell^x + (y'-y)d_\ell^y\right]^2 + 4(Z_0-z)^2 (d_\ell^z)^2}{4R_0^3},
\end{multline}
where $R_0$ is the distance between the virtual monostatic element located at $(x',y',Z_0)$ and the point scatterer at $(x,y,z)$, expressed as
\begin{equation}
\label{eq:R0}
    R_0 = \left[ (x'-x)^2 + (y'-y)^2 + (Z_0-z)^2 \right]^{\frac{1}{2}}.
\end{equation}

Centering the target to the origin of the $(x,y,z)$ coordinate system and considering $(x'-x),(y'-y) \ll Z_0$, we can acquire the improved approximation of the round-trip distance between the $\ell$-th Tx/Rx pair and the point scatterer as
\begin{equation}
\label{eq:Rl_best_approximation}
    R_\ell^{RT}= R_\ell^T + R_\ell^R \approx 2 R_0 + 2 d_\ell^z + \frac{(d_\ell^x)^2 + (d_\ell^y)^2}{4 Z_0}.
\end{equation}

\subsection{Signal Model}
\label{subsec:signal_model}
Consider the multi-planar multistatic array whose Tx and Rx elements are located at $(x_T,y_T,z_\ell)$ and $(x_R,y_R,z_\ell)$, respectively, and a distributed target occupying the volume $V$ at locations $(x,y,z)$ in \mbox{3-D} space with a continuous reflectivity function given by $o(x,y,z)$.
Without loss of generality, under the Born approximation for the scattering process and an isotropic antenna assumption, the received signal can be written as
\begin{equation}
    \label{eq:received_general}
    s(x_T,x_R,y_T,y_R,z_\ell,t) = \iiint_V \frac{o(x,y,z)}{R_\ell^T R_\ell^R} p \left( t - \frac{R_\ell^T}{c} - \frac{R_\ell^R}{c} \right) dx dy dz,
\end{equation}
where $p(t)$ is the transmitted signal, $t$ is the fast-time variable, $c$ is the speed of the EM wave, and $R_\ell^T$, $R_\ell^R$ are given in (\ref{eq:Rl_of_xT_xR_yT_yR}).

The Fourier transform of (\ref{eq:received_general}) with respect to $t$ yields the frequency spectrum as
\begin{equation}
    \label{eq:received_k}
    s(x_T,x_R,y_T,y_R,z_\ell,f) = P(f) \iiint_V \frac{o(x,y,z)}{R_\ell^T R_\ell^R} e^{-jk(R_\ell^T + R_\ell^R)} dx dy dz,
\end{equation}
where $k = 2\pi f/ c$ is the instantaneous wavenumber and $P(f)$ is the Fourier transform of $p(t)$. 
Image recovery requires the inversion of (\ref{eq:received_k}) to produce $o(x,y,z)$; however, given arbitrary sampling locations, the image cannot be computed efficiently using existing techniques \cite{alvarez2019freehand,alvarez2021freehand,alvarez2021freehandsystem,alvarez2021system,alvarez2021towards,garcia20203DSARProcessing,wu2020multilayered}.
The frequency-domain model of the received signal (\ref{eq:received_k}) is valid for any UWB radar signaling scheme, including frequency-modulated continuous-wave (FMCW), phase-modulated continuous wave (PMCW), and orthogonal frequency-division multiplexing (OFDM), which is commonly employed in 5G and IoT applications \cite{roos2019radar}. 
Furthermore, prior research on freehand imaging and similar IoT applications has employed a purely stepped-frequency FMCW signal model \cite{alvarez2019freehand,alvarez2021freehand,alvarez2021freehandsystem,alvarez2021system}. 
Similarly, Google Pixel 4 utilizes a Google Soli 60 GHz mmWave FMCW radar for sensing \cite{basrawi2021reverse}. 

However, the derivation of (\ref{eq:Rl_best_approximation}) enables efficient compensation of multistatic multi-planar data by careful handling of the phase.
To achieve the proposed compensation, we express the frequency response of the virtual planar monostatic array, whose elements are located at $(x',y',Z_0)$, as
\begin{equation}
    \label{eq:virtual_planar}
    \hat{s}(x',y',f) = P(f) \iiint_V \frac{o(x,y,z)}{R_0^2} e^{-j2kR_0} dx dy dz,
\end{equation}
where $R_0$ is given by (\ref{eq:R0}), $x'$ and $y'$ are the midpoints between each Tx/Rx pair, and $Z_0$ is the plane on which the samples are projected. 
From the analysis in Section \ref{subsec:virtual_array}, the relationship between the multi-planar multistatic response and the virtual monostatic array response is given by
\begin{equation}
    \label{eq:multiplanar_compensation}
    \hat{s}(x',y',f) \approx s(x_T,x_R,y_T,y_R,z_\ell,f) e^{j k\beta_\ell},
\end{equation}
where
\begin{equation}
    \beta_\ell = 2 d_\ell^z + \frac{(d_\ell^x)^2 + (d_\ell^y)^2}{4 Z_0},
\end{equation}
is the near-field residual phase term due to the arbitrary scanning and MIMO effects in the near-field, as derived in (\ref{eq:Rl_best_approximation}). 
Hence, the virtual planar monostatic response can be efficiently acquired from the irregular samples by removing the residual phase to simultaneously account for the multi-planar scanning geometry and near-field multistatic effects. 
The novel phase compensation technique derived in this section efficiently reduces the dimensionality of the MIMO-SAR imaging problem and projects multi-planar samples onto a single plane to enable computationally tractable algorithms for image reconstruction.

\section{Efficient Image Reconstruction Algorithms for Near-Field SAR}
\label{sec:image_reconstruction}
In this section, we review traditional planar SAR image reconstruction methods using efficient Fourier-based solutions to recover EM images \cite{smith2022ThzToolbox} and propose a novel technique for multi-planar multistatic SAR. 
Existing research on irregularly sampled SAR imaging problems employs the gold-standard back-projection algorithm (BPA) \cite{alvarez2019freehand,alvarez2021freehand,alvarez2021freehandsystem,alvarez2021system,alvarez2021towards,garcia20203DSARProcessing,wu2020multilayered}.
However, this approach is computationally infeasible for most edge and mobile applications.
To overcome this challenge, we employ the approximation in (\ref{eq:multiplanar_compensation}) to project multi-planar data to a planar-sampled scenario to satisfy the requirements for efficient image reconstruction.
The Fourier-based algorithm detailed in the subsequent analysis is known as the range migration algorithm (RMA) or \mbox{$f$-$k$} algorithm, and has been explored in greater detail elsewhere \cite{sheen2001three,yanik2019cascaded,smith2022ThzToolbox,gao2018_1D_MIMO,fan2020linearMIMOArbitraryTopologies}.

The key step to efficiently invert the integral in (\ref{eq:virtual_planar}) is to represent the spherical wave term as a superposition of plane waves using the method of stationary phase (MSP) \cite{smith2022ThzToolbox,yanik2019sparse}, such that
\begin{equation}
\label{eq:MSP_expansion}
    \frac{e^{-j2 k R_0}}{R_0} \approx \iint_A \frac{e^{-j(k_x'(x-x') + k_y'(y-y') + k_z(z-Z_0))}}{k_z} dk_x' dk_y',
\end{equation}
where
\begin{equation}
    k_z^2 = 4k^2 - (k_x')^2 - (k_y')^2,
\end{equation}
and $A$ is the region in $k_x'$-$k_y'$ space occupied by the spherical wavefront.

Following the analysis in \cite{smith2022ThzToolbox,yanik2020development}, substituting (\ref{eq:MSP_expansion}) into (\ref{eq:virtual_planar}) and rearranging the phase terms to leverage the Fourier relationships yields
\begin{equation}
    O(k_x,k_y,k_z) = \hat{S}(k_x',k_y',f) \frac{k_z}{P(f)} e^{-j k_z Z_0},
\end{equation}
where $O(k_x,k_y,k_z)$ and $\hat{S}(k_x',k_y',k)$ are the spatial spectral representations of the reflectivity function $o(\cdot)$ and the array response $\hat{s}(\cdot)$, respectively.
Because the primed and unprimed coordinate systems are coincident, the distinction can be dropped for the remaining analysis.
Hence, the RMA image recovery process can be summarized as
\begin{equation}
\label{eq:RMA_final}
    o(x,y,z) = \text{IFT}_{\text{3D}}^{(k_x,k_y,k_z)}\left[ \mathcal{S} \left[ \text{FT}_{\text{2D}}^{(x',y')} \left[ \hat{s}(x',y',f) \right] \frac{k_z}{P(f)} e^{-j k_z Z_0} \right] \right],
\end{equation}
where $\text{FT}[\cdot]$ and $\text{IFT}[\cdot]$ are the forward and inverse Fourier transform operators, respectively; $\mathcal{S}$ is the Stolt interpolation operator required to compensate for the spherical wavefront \cite{smith2022ThzToolbox}; and $\hat{s}(\cdot)$ is obtained from (\ref{eq:multiplanar_compensation}).
The spatial resolution along each dimension of the recovered image is given by
\begin{align}
\label{eq:spatial_resolution}
\begin{split}
    \delta_x &= \frac{\lambda_c Z_0}{2 D_x}, \\
    \delta_y &= \frac{\lambda_c Z_0}{2 D_y}, \\
    \delta_z &= \frac{c}{2B},
\end{split}
\end{align}
where $D_x$ and $D_y$ are the sizes of the aperture along the $x$- and $y$-directions, respectively; $B$ is the system bandwidth; and $\lambda_c$ is the wavelength of the center frequency \cite{smith2022ThzToolbox,yanik2019sparse,gao2018_1D_MIMO}.

While (\ref{eq:RMA_final}) provides an efficient solution for planar array imaging problems, its application to irregular scanning geometries requires a discussion of several key issues.
Applying the compensation technique in (\ref{eq:multiplanar_compensation}) for irregularly sampled data, the multi-planar data can be approximately projected to planar sampling; however, they are likely non-uniform at the positions $(x',y',Z_0)$ along the $x$- and $y$-directions.
Traditional efficient implementations rely on the common fast Fourier transform (FFT) algorithm, but recent work on non-uniform planar MIMO-SAR \cite{gao2018_1D_MIMO,fan2020linearMIMOArbitraryTopologies} and irregular MIMO real aperture radar (MIMO-RAR) \cite{wang20203} imaging has produced solutions using a non-uniform FFT (NUFFT) approach employing fast Gaussian gridding (FGG), as discussed in \cite{greengard2006nufft}, for the Fourier transforms and Stolt interpolation step in (\ref{eq:RMA_final}). 
The sampling criteria for the nonuniform planar case are discussed in detail in \cite{gao2018_1D_MIMO,fan2020linearMIMOArbitraryTopologies,wang20203} and apply correspondingly to irregular scanning scenarios after multi-planar compensation. 
Similarly, the FGG-NUFFT technique is employed in this study to perform the proposed RMA efficiently on irregularly sampled planar data.

For the multi-planar sampling scenario discussed in Section \ref{subsec:multiplanar_SAR}, the RMA cannot be applied directly without multi-planar compensation because the data are sampled on different $z$-planes, as discussed in Section \ref{sec:results}.
If the RMA is applied to the raw multi-planar data, the forward Fourier transform in (\ref{eq:RMA_final}) is invalid because the data along the $x'$ and $y'$-directions are not coplanar and the resulting image will suffer from significant distortion, rendering the resulting images unusable in most cases.

Sampling considerations for image reconstruction remain identical to those in analyses elsewhere \cite{yanik2019sparse,sheen2001three}, after the multi-planar compensation algorithm.
Baseband frequency sampling criteria can be determined by the maximum range for a given application.
As given in \cite{sheen2010near}, the maximum frequency sampling interval is given by $\Delta_f < c/(2R_\text{max})$, where $R_\text{max}$ is the maximum target range. 
Although spatial sampling criteria are not guaranteed for irregular SAR scanning, if the relationship between the capture rate of the radar and the velocity of the radar platform is tuned appropriately during system design, undersampling artifacts are typically minimal \cite{alvarez2019freehand,alvarez2021freehand,alvarez2021freehandsystem,alvarez2021system}. 
To avoid spatial undersampling, the lower bound of the pulse repetition frequency (PRF) can be computed using $\text{PRF} > 4v_\text{max}/\lambda_c$, where $v_\text{max}$ is the maximum velocity for a certain application. 
For example, assuming that the maximum velocity of the human hand for a freehand SAR is 1 m/s and a center frequency of 79 GHz, the lower bound of the PRF is approximately 1.06 kHz. 
It is important to note that the number of captures increases proportionally to the PRF; hence, at high velocities, a large number of samples are captured. 
The computational performance of traditional techniques that employ the BPA degrades substantially when many samples are captured. 
On the other hand, the signal-to-noise ratio can be improved by increasing the number of samples at the cost of an increased computational burden.
Hence, an efficient algorithm for multi-planar MIMO-SAR imaging is required to enable many such technologies. 

In terms of computational complexity, our proposed algorithm offers a significant advantage over existing techniques in the literature \cite{alvarez2019freehand,alvarez2021freehand,alvarez2021freehandsystem,alvarez2021system,alvarez2021towards,garcia20203DSARProcessing,wu2020multilayered}, which employ the BPA, whose computational complexity is on the order of $O(N^6)$ \cite{yanik2020development,fan2020linearMIMOArbitraryTopologies}. 
The time complexity of the RMA and its FGG-NUFFT variants are investigated in the literature \cite{fan2020linearMIMOArbitraryTopologies,wang20203} and the multi-planar compensation step proposed in this article presents negligible computational expense to the RMA, which is on the order of $O(N^3 \log{N})$ \cite{lopez20003,sheen2001three}.
Hence, as discussed in Section \ref{sec:results}, the algorithm proposed in this article offers comparable imaging performance to the BPA with tractable execution time for mobile platforms, similar to the RMA.

\begin{figure}[h]
    \centering
    \includegraphics[width=0.35\textwidth]{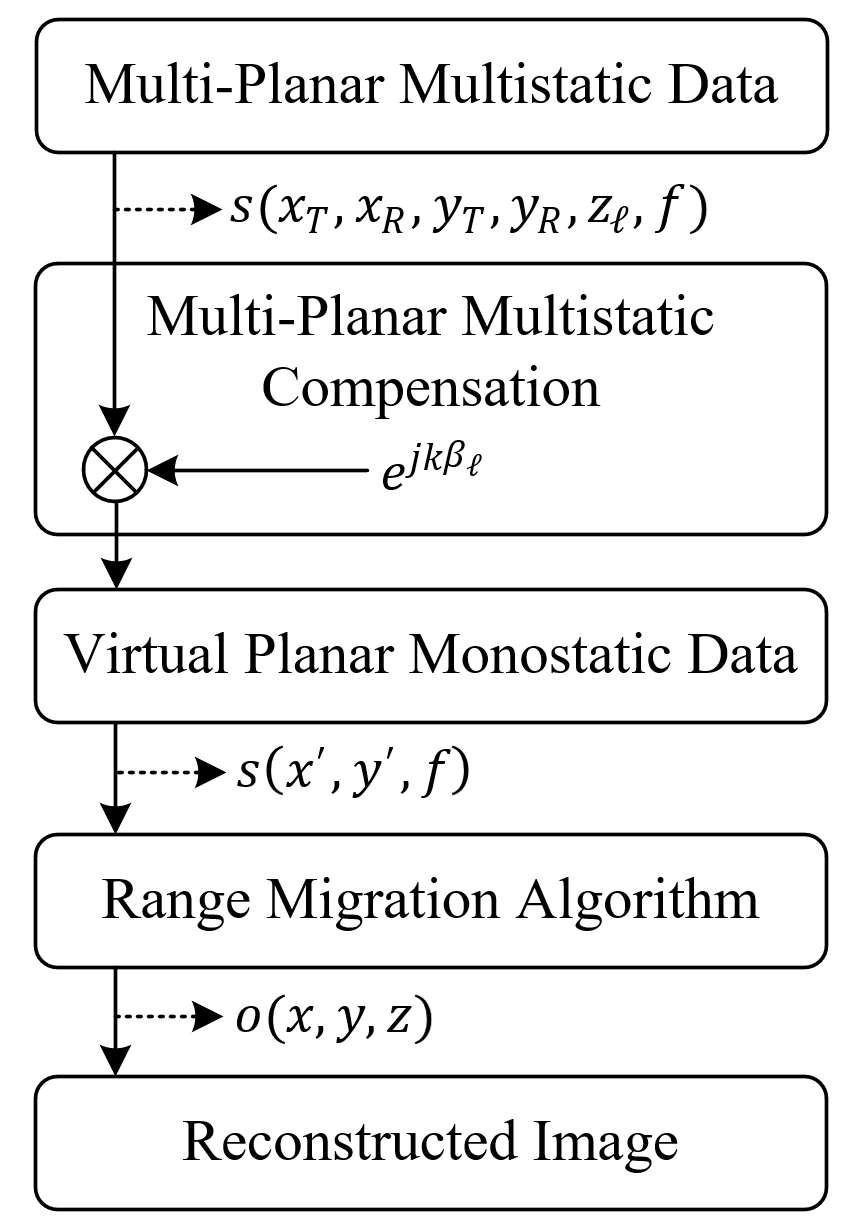}
    \caption{The complete image reconstruction process from irregular sampling compensation to RMA image recovery.}
    \label{fig:image_reconstruction_process_complete}
\end{figure} 

The novel enhanced reconstruction process for efficient near-field SAR imaging with irregular scanning geometries is summarized in Fig. \ref{fig:image_reconstruction_process_complete}.
Using the analysis in Section \ref{sec:system_model}, irregular scanning geometries can be modeled as multi-planar sampling scenarios, as shown in Fig. \ref{fig:multiplanar}, and compensated by removing the residual phase due to the multi-planar multistatic conditions. 
The key difference between the traditional RMA and the proposed algorithm is the alignment of the multi-planar multistatic (MIMO-SAR) data to virtual planar monostatic data. 
This crucial step compensates for both the sampling irregularities and multistatic MIMO effects simultaneously, while significantly reducing the dimensionality, from \mbox{6-D} $(x_T,x_R,y_T,y_R,z_\ell,f)$ to \mbox{3-D} $(x',y',f)$, and subsequently the computational complexity. 
Finally, virtual planar monostatic data are used to efficiently recover the image using the RMA.
In simulation and empirical studies on irregular SAR scanning geometries, the proposed algorithm is applied to efficiently produce high-resolution \mbox{3-D} images previously infeasible due to algorithmic deficiencies. 

\begin{figure*}[ht]
    \centering
    \includegraphics[width=0.85\textwidth]{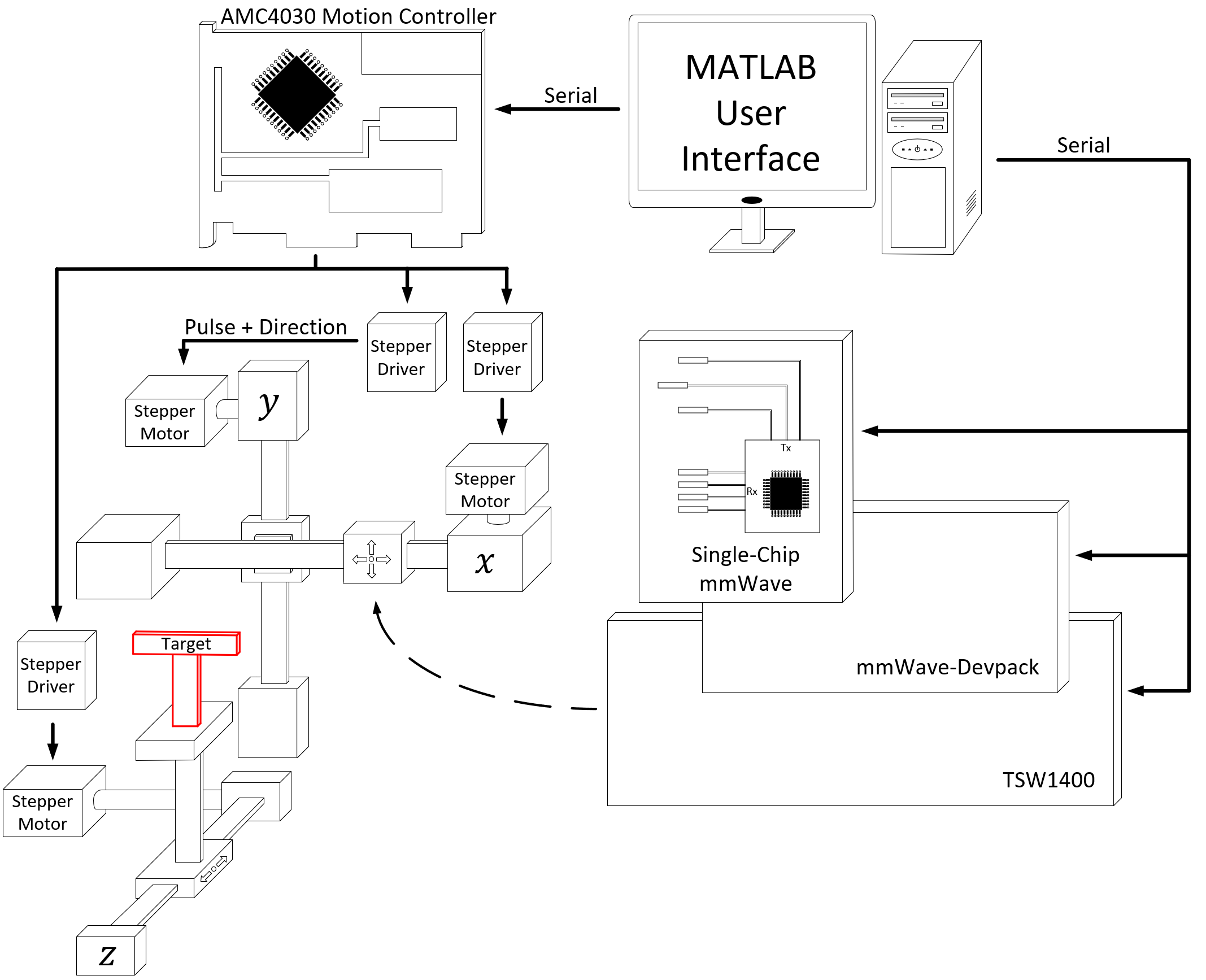}
    \caption{System design for \mbox{3-D} scanner with radar mounted on planar $x$-$y$ rails and target mounted on a linear $z$ rail. The TI radar, data capture card, and mechanical scanner are controlled by MATLAB via USB serial interface.}
    \label{fig:scanner_xyz}
\end{figure*}

\section{Multi-Planar Multistatic Imaging Hardware Prototype}
\label{sec:prototype}
In this section, we discuss the hardware prototype implementation for empirically validating the proposed imaging algorithm by collecting multi-planar multistatic SAR data.
The hardware architecture of the mmWave imaging system is illustrated in Fig. \ref{fig:scanner_xyz}.
A Texas Instruments (TI) mmWave MIMO radar is mounted on an $x$-$y$ planar scanner. 
The TI AWR1443BOOST radar with a bandwidth of \mbox{4 GHz} from \mbox{77 GHz} to \mbox{81 GHz} is mounted on a TI mmWave-Devpack and TSW1400 data capture card to store the data from the SAR scan and transfer it to the PC, where the image recovery algorithm is implemented in MATLAB.
The TI AWR1443BOOST is equipped with a MIMO array consisting of two Tx elements spaced by $2\lambda_c$ and four Rx elements spaced by $\lambda_c/2$ \cite{yanik2019sparse}.
Although 5G and IoT applications commonly employ an OFDM modulation scheme, the TI radar employed for the following experiments utilizes FMCW signaling. 
However, FMCW radar has been utilized for smartphone applications, notably the Google Pixel 4, which is equipped with a Google Soli FMCW radar \cite{basrawi2021reverse}. 
The proposed range migration-based algorithm is applicable to both OFDM and FMCW radars; hence, the results discussed in the following section are relevant for a wide array of 5G, IoT, smartphone, and automotive applications \cite{zhang2015ofdm}.

Additionally, a linear rail is used to move the target along the $z$-direction to collect multi-planar multistatic data under the geometry discussed in Section \ref{subsec:multiplanar_SAR}.
All three $x$-$y$-$z$ rails are driven by stepper motors controlled by an AMC4030 motion controller, and the scanning process and radar set up are handled in MATLAB.
Additional details on system development and device calibration can be found in \cite{yanik2020development}.
The images are reconstructed using MATLAB implementations running on a desktop PC equipped with a 12-core AMD Ryzen 9 3900X running at 4.6 GHz with 64 GB of memory.
Using this hardware prototype, data can be collected for many target scenarios under the multi-planar multistatic scenario by performing multiple planar SAR scans with the target at different $z$-locations.
To emulate irregular scanning geometries, the data collected throughout $x$-$y$-$z$ space are subsampled, as discussed in Section \ref{sec:results}.
Implementations in the literature employ multi-camera infrared camera systems to track the radar as it is moved through space by the user \cite{alvarez2019freehand,alvarez2021freehand}.
Other studies on irregular scanning geometries have explored freehand imaging using a stereo camera with an IMU for positioning estimation \cite{alvarez2021towards} and UAV near-field imaging with a laser rangefinder and a real-time kinematic (RTK) system for localization \cite{garcia20203DSARProcessing}. 
These implementations, among others \cite{alvarez2021freehandsystem,alvarez2021system,wu2020multilayered,kan2020automotiveSAR}, demonstrate the viability of high-resolution sensors for precise positioning to enable novel imaging techniques using UWB mmWave radars.
Hence, this study focuses on improving the computational efficiency of the imaging technique and assumes that the radar position is known across an irregularly sampled geometry.

\section{Measurement and Imaging Results}
\label{sec:results}
In this section, we validate the proposed algorithm derived in Sections \ref{sec:system_model} and \ref{sec:image_reconstruction}, as illustrated in Fig. \ref{fig:image_reconstruction_process_complete}.
Irregular scanning geometries were simulated using the simulation platform developed in \cite{smith2022ThzToolbox}, and image reconstruction results are shown by comparing our enhanced method to the gold standard BPA and RMA without multi-planar multistatic compensation.
Similarly, using the \mbox{3-D} mechanical system detailed in Fig. \ref{fig:scanner_xyz}, irregular scanning geometries were emulated by collecting planar scans with the target at different $z$-locations and subsampling the collected data.
Imaging results, comparing the proposed algorithm with the BPA and RMA, demonstrate the computational advantage of our technique while achieving nearly identical spatial resolution.
The proposed multi-planar multistatic compensation algorithm achieves image quality comparable to that of the BPA while offering time and space complexity on par with the RMA.

\begin{figure}[h]
    \begin{subfigure}[b]{0.48\textwidth}
         \centering
         \includegraphics[width=\textwidth]{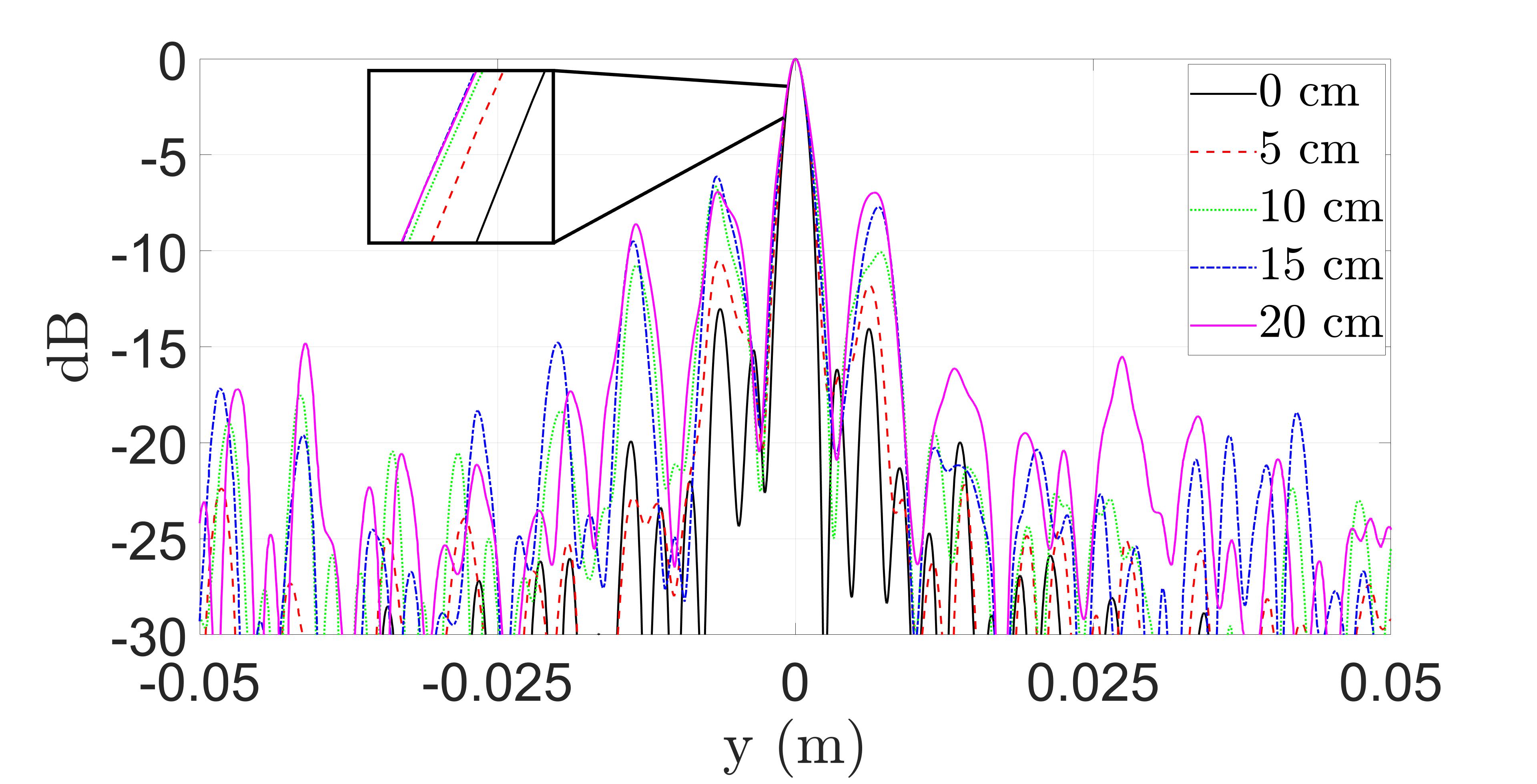}
         \caption{}
         \label{fig:sim0_PSF_y}
    \end{subfigure}
    \begin{subfigure}[b]{0.48\textwidth}
         \centering
         \includegraphics[width=\textwidth]{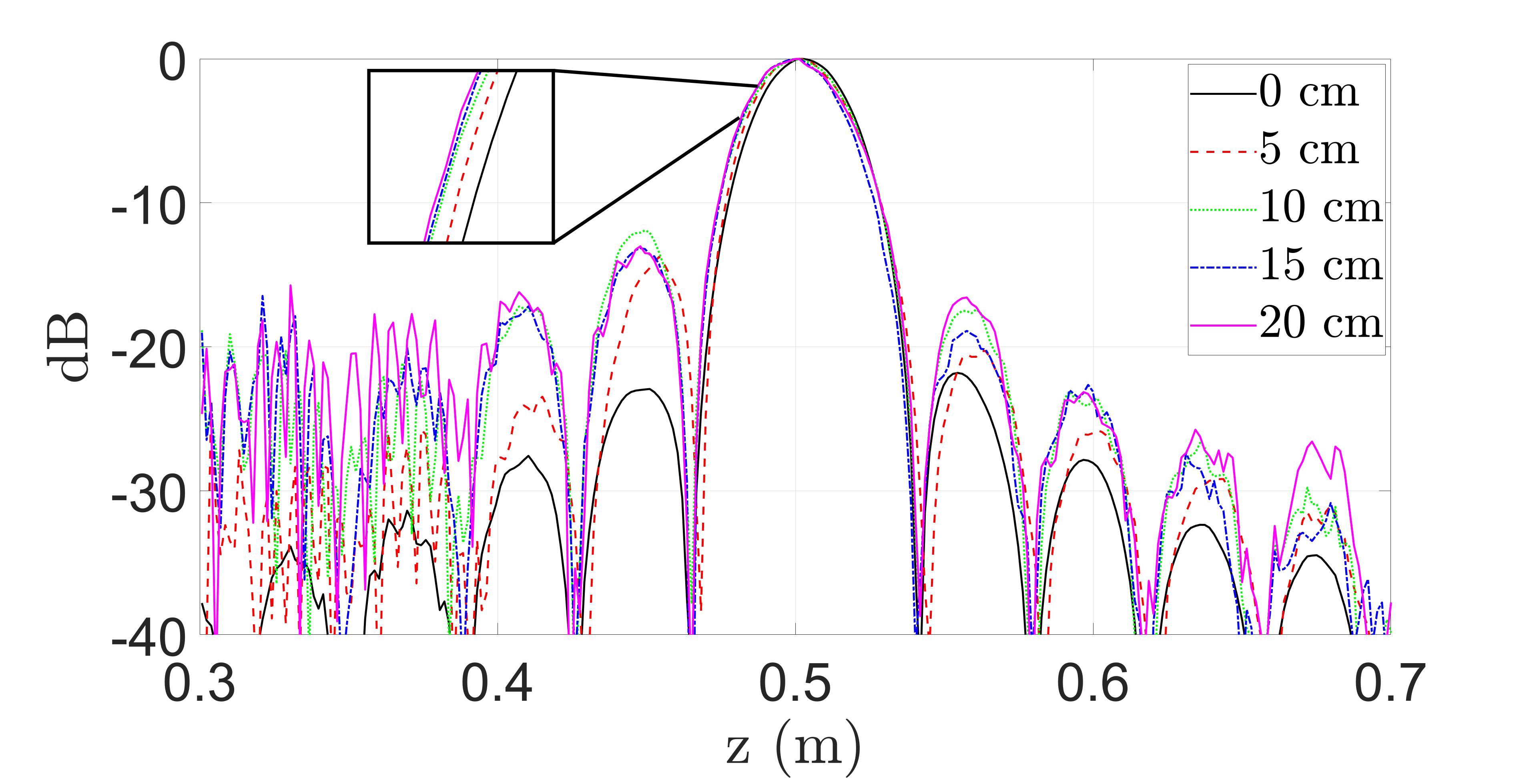}
         \caption{}
         \label{fig:sim0_PSF_z}
    \end{subfigure}
\caption{Comparison of point spread function (PSF) resolution along the (a) $y$- and (b) $z$-direction with varying maximum distance from the reference plane, $Z_0$, to the samples at $z_\ell$. The distance between the samples and reference plane, $\Delta_z^\text{max}$, is varied from \mbox{0 cm} (linear) to \mbox{20 cm} with a step size of \mbox{5 cm}. The linear case (0 cm) is computed with the conventional RMA. Each of the remaining PSFs are computed using our proposed algorithm.}
\label{fig:sim0_PSF}
\end{figure}

\subsection{Simulated SAR Imaging Results}
\label{subsec:sim}
\begin{figure}[h]
\centering
    \begin{subfigure}[b]{0.35\textwidth}
         \centering
         \includegraphics[width=\textwidth]{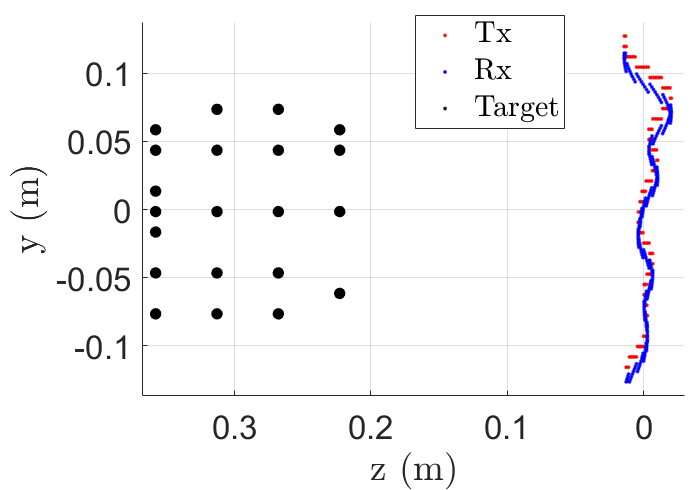}
         \caption{}
         \label{fig:sim1_UTD_scenario}
    \end{subfigure}
    \begin{subfigure}[b]{0.35\textwidth}
         \centering
         \includegraphics[width=\textwidth]{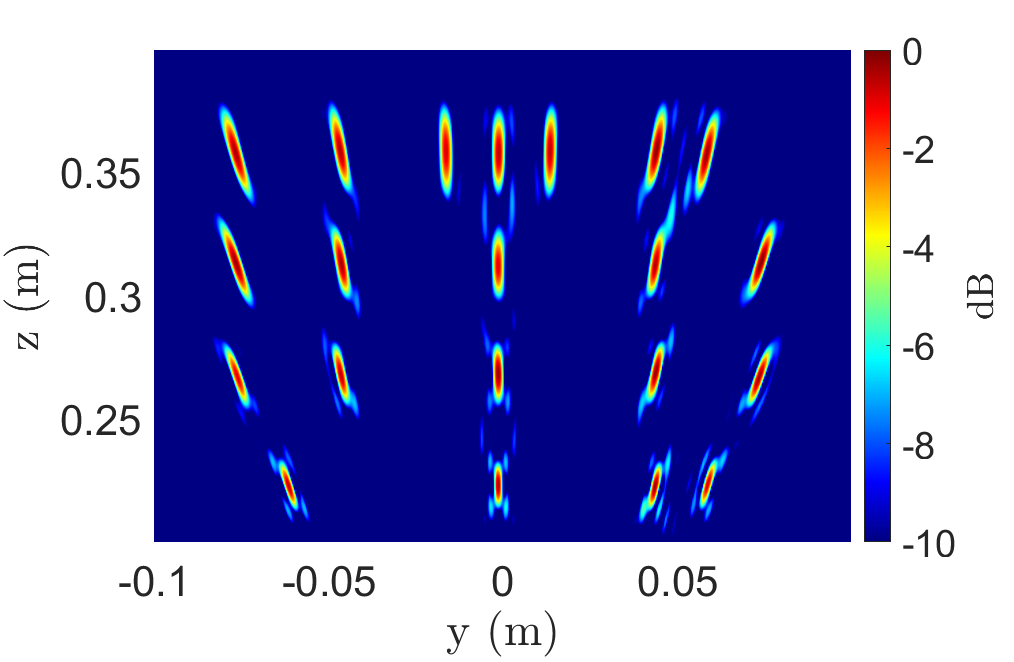}
         \caption{}
         \label{fig:sim1_UTD_BPA}
    \end{subfigure}
    \vskip\baselineskip
    \begin{subfigure}[b]{0.35\textwidth}
         \centering
         \includegraphics[width=\textwidth]{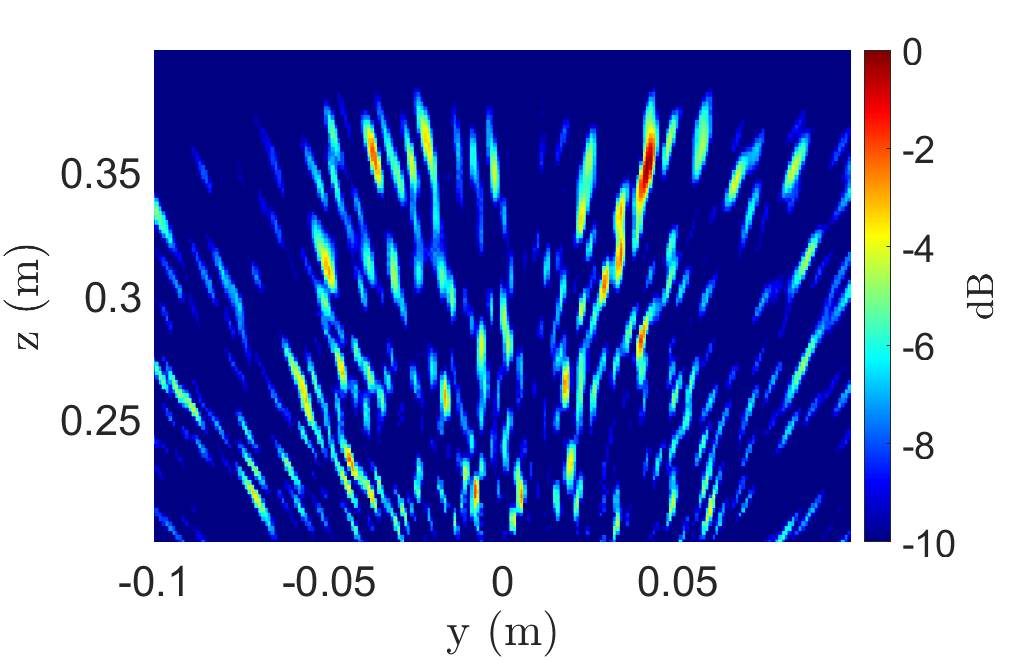}
         \caption{}
         \label{fig:sim1_UTD_RMA}
    \end{subfigure}
    \begin{subfigure}[b]{0.35\textwidth}
         \centering
         \includegraphics[width=\textwidth]{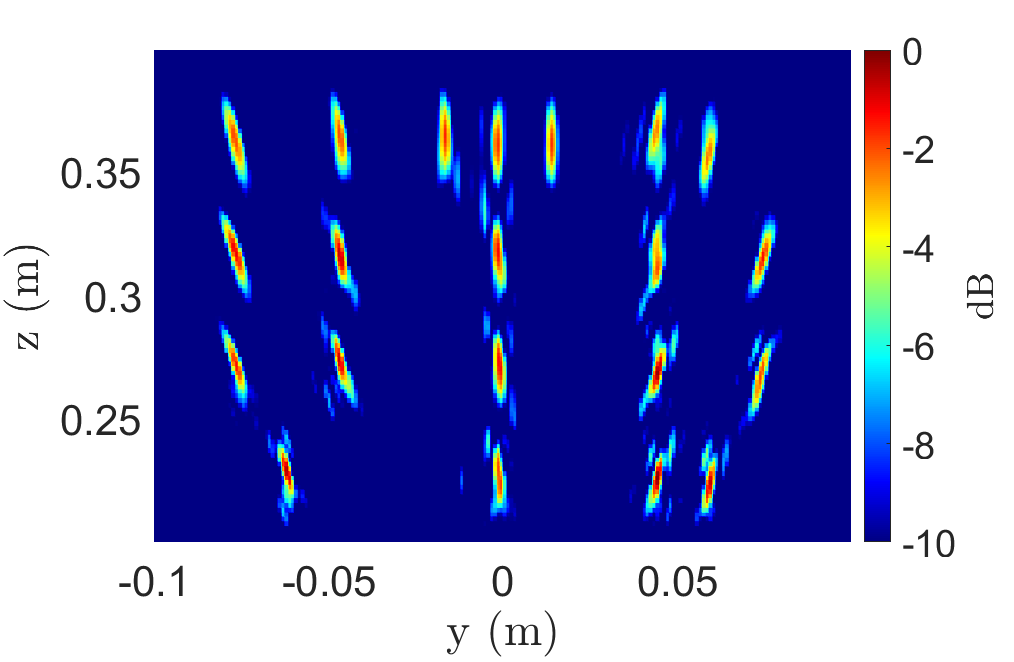}
         \caption{}
         \label{fig:sim1_UTD_RMA_FFH}
    \end{subfigure}
\caption{(a) Irregular scanning geometry for ``UTD'' scenario consisting of a multi-linear array in the $y$-direction at \mbox{$x$ = 0 m} and corresponding imaging results using the (b) BPA \mbox{(296.3 s)}, (c) RMA without multi-planar multistatic compensation \mbox{(29 ms)}, and (d) our proposed algorithm \mbox{(30 ms)}.}
\label{fig:sim1_UTD}
\end{figure}

To validate our proposed algorithm in simulation, we consider three distinct scenarios.
First, we investigate the impact of array irregularities on image resolution.
We consider the point spread functions (PSFs) of several multi-planar MIMO-SAR scenarios and compare them with an ideal planar scanning scenario to analyze the range and cross-range resolution of the proposed algorithm. 
We assume a single, ideal point target located at \mbox{$(0, 0, 0.5$ m$)$} in \mbox{3-D} space for the PSF simulation. 
For comparison, an ideal linear MIMO-SAR pattern is generated along with several irregular SAR scanning patterns with increasing irregularity.
Each non-cooperative motion track is generated by a semi-smooth, random curve spanning \mbox{$y' \in [-12.5, 12.5]$ cm} with varying $z_\ell$ around \mbox{$Z_0 = 0$ m} with 256 sampling locations, as shown in Fig. \ref{fig:sim1_UTD_scenario}. 
To analyze the impact of $d_\ell^z = z_\ell - Z_0$, the distance between the reference plane at $Z_0$, and the samples at $z_\ell$, we simulate several multi-planar multistatic with increasing variance of $d_\ell^z$, as shown in Fig. \ref{fig:sim0_PSF}. 
The absolute maximum distance, $\Delta_z^\text{max} \triangleq \max |d_\ell^z|$, varies from \mbox{0 cm}, the linear case, to \mbox{20 cm} with a step size of \mbox{5 cm}. 
At a center frequency of \mbox{79 GHz}, \mbox{$\Delta_z^\text{max} =$ 20 cm} is more than 50 times the wavelength, $\lambda_c =$ \mbox{3.79 mm}. 
Prior work on freehand smartphone imaging system design assumes deviations on the order of several centimeters \cite{alvarez2021towards}. 
Along the cross-range dimension, which is symmetric along both the $x$- and $y$-directions, the resolution is minimally affected by the irregular scanning geometry when the algorithm is applied, as shown in Fig. \ref{fig:sim0_PSF_y}. 
However, a direct relationship between $\Delta_z^\text{max}$ and the main beamwidth is observed along with decreased sidelobe suppression compared to the ideal linear case, where the traditional RMA can be employed directly. 
Along the $z$-direction, the resolution of the proposed algorithm suffers as $\Delta_z^\text{max}$ increases, but remains quite similar to the resolution of the linear case, as shown in Fig. \ref{fig:sim0_PSF_z}. 
Hence, our efficient algorithm achieves a focusing performance comparable to that of the ideal linear or planar RMA case while allowing for irregular scanning geometries with large deviations from the reference plane in the $z$-direction. 

To evaluate the performance of the algorithm for more complex targets, a linear array along the $y$-axis is simulated as shown in Fig. \ref{fig:sim1_UTD_scenario} with 21 point scatterers arranged as the letters ``UTD.''
Again, irregular array locations are generated by a semi-smooth, random curve spanning \mbox{$y' \in [-12.5, 12.5]$ cm} and \mbox{$z_\ell \in [-2.5, 2.5]$ cm} with 256 sampling locations.
As shown in Fig. \ref{fig:sim1_UTD_BPA}, the gold standard BPA recovers each point scatterer without artifacts; however, computing the BPA image requires 296.3 s on our machine. 
The RMA without multi-planar multistatic compensation and the proposed algorithm are considerably more efficient, requiring only 30 ms for computation.
However, while the RMA image in Fig. \ref{fig:sim1_UTD_RMA} is significantly distorted and the image is lost, our proposed method resolves the point targets comparably to the BPA and requires a fraction of the computation time.

\begin{figure}[h]
    \centering
    \includegraphics[width=0.5\textwidth]{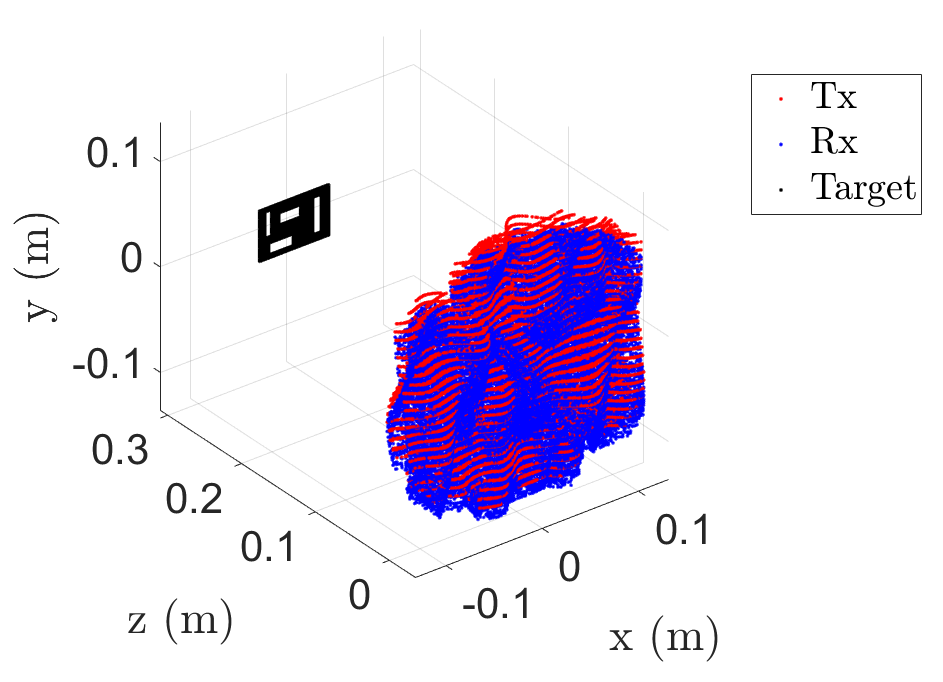}
    \caption{Irregular scanning geometry for cutout consisting of a multi-planar array along the $x$- and $y$-directions.}
    \label{fig:sim2_cutout2_scenario}
\end{figure}

\begin{figure}[h]
\centering
    \begin{subfigure}[b]{0.35\textwidth}
         \centering
         \includegraphics[width=\textwidth]{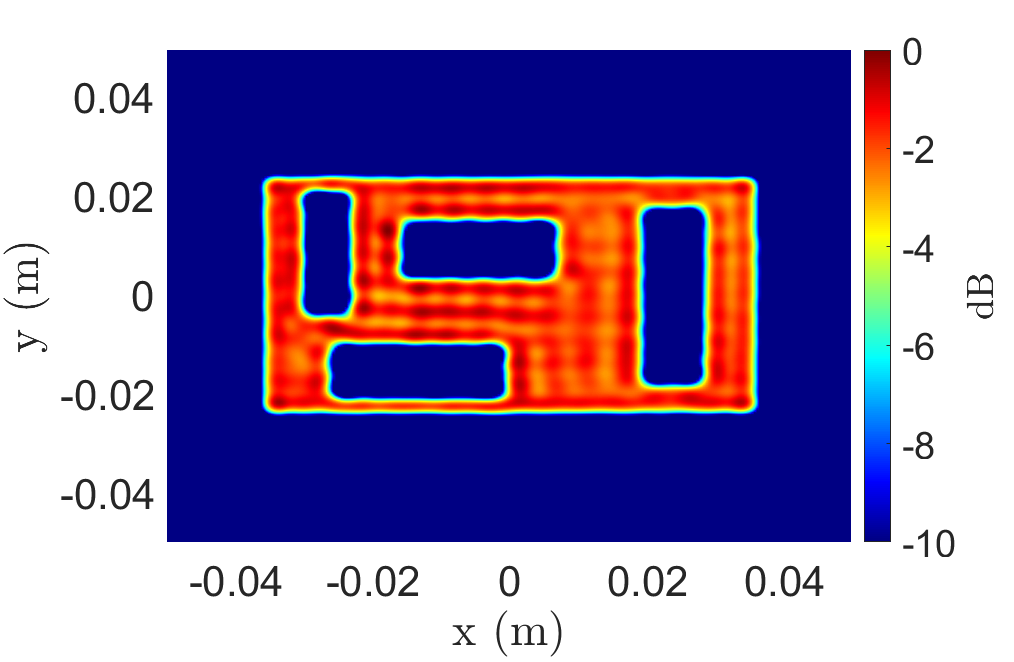}
         \caption{}
         \label{fig:sim2_cutout2_BPA}
    \end{subfigure}
    \begin{subfigure}[b]{0.35\textwidth}
         \centering
         \includegraphics[width=\textwidth]{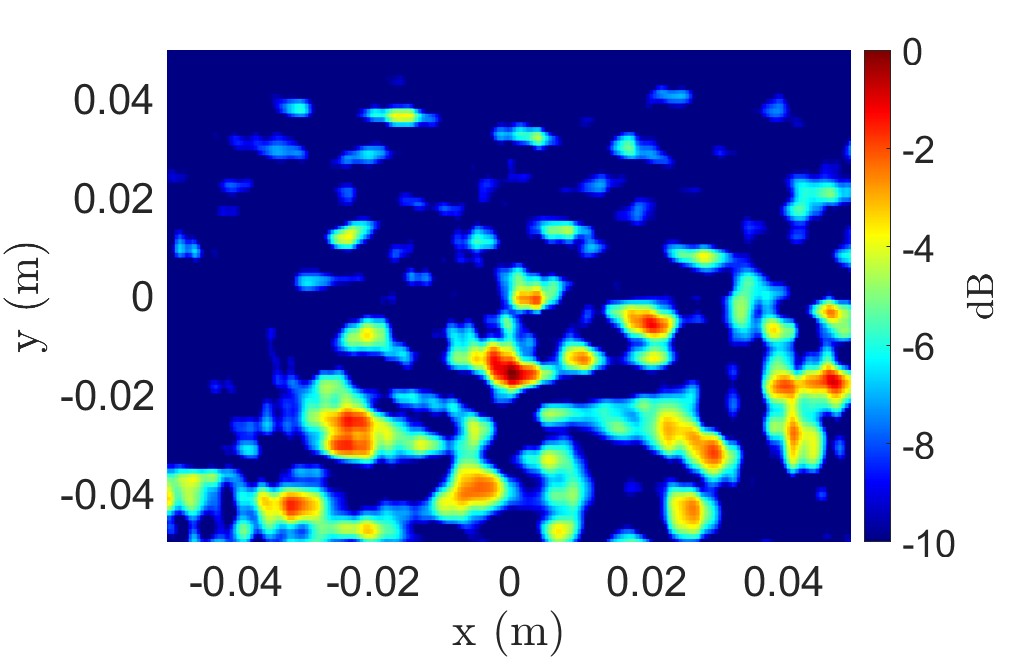}
         \caption{}
         \label{fig:sim2_cutout2_RMA}
    \end{subfigure}
    \vskip\baselineskip
    \begin{subfigure}[b]{0.35\textwidth}
         \centering
         \includegraphics[width=\textwidth]{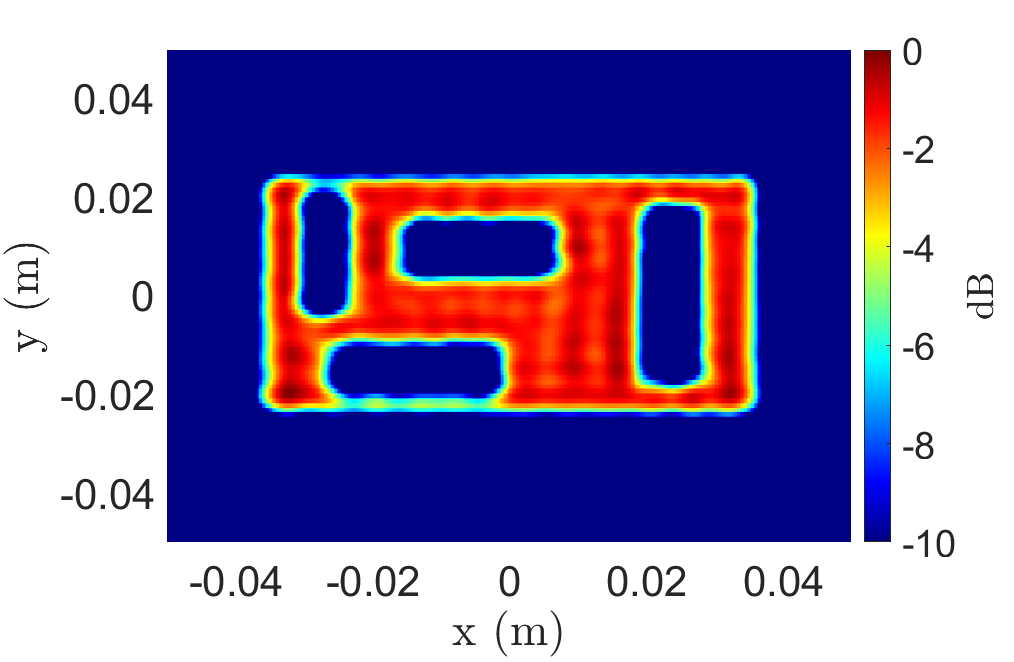}
         \caption{}
         \label{fig:sim2_cutout2_RMA_FFH}
    \end{subfigure}
    \begin{subfigure}[b]{0.35\textwidth}
         \centering
         \includegraphics[width=\textwidth]{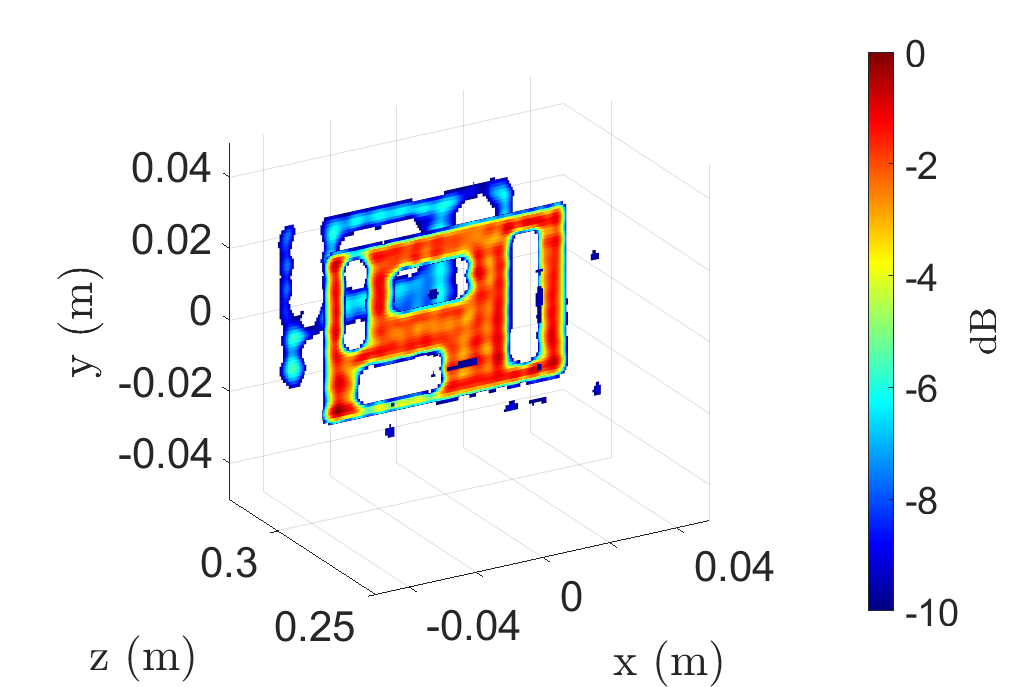}
         \caption{}
         \label{fig:sim2_cutout2_RMA_FFH_3D}
    \end{subfigure}
\caption{Imaging results for the scenario in Fig. \ref{fig:sim2_cutout2_scenario} using the (a) BPA \mbox{(1324.8 s)}, (b) RMA without multi-planar multistatic compensation \mbox{(1.1 s)}, (c) proposed algorithm at the \mbox{$z$ = 300 mm} plane \mbox{(1.1 s)}, and (d) the \mbox{3-D} reconstructed image using the proposed technique \mbox{(4.8 s)}.}
\label{fig:sim2_cutout2}
\end{figure}

Considering the more broadly applicable \mbox{2-D} scanning case, a \mbox{2-D} multi-planar multistatic scenario is simulated with a solid target located at \mbox{$z$ = 300 mm}, as shown in Fig. \ref{fig:sim2_cutout2_scenario}.
The target is a rectangular strip with cutouts of various sizes and the irregular sampling geometry is generated as a \mbox{2-D} semi-smooth random curve occupying \mbox{$x' \in [-12.5, 12.5]$ cm}, \mbox{$y' \in [-12.5, 12.5]$ cm}, and \mbox{$z_\ell \in [-2.5, 2.5]$ cm} with 102956 sampling locations. 
The number of sampling locations is selected to approximate a virtual array spanning $[-12.5, 12.5]$ cm along the $x$- and $y$-directions, a realistic aperture size for many applications, while satisfying the sampling condition, such that the distance between subsequent sampling points is always less than $\lambda_c/4$.
Since the target is located on a single $z$-plane parallel to the planar projection after our compensation technique, a \mbox{2-D} $x$-$y$ image is recovered at \mbox{$z$ = 300 mm}. 
Again, while the BPA yields a robust reconstruction, the computation time is excessive for most applications, requiring 1324.8 s on a desktop machine.
On the other hand, the proposed algorithm outperforms the RMA significantly in terms of image quality, nearly matching that of the BPA with only slight artifacting, while demonstrating superior efficiency to the BPA computing a high-resolution \mbox{2-D} image in only 1.1 s.
Similarly, \mbox{3-D} images can be reconstructed using these methods.
The \mbox{3-D} reconstructed image using the proposed algorithm is shown in Fig. \ref{fig:sim2_cutout2_RMA_FFH_3D}, requiring 4.8 s to compute, while the RMA and BPA are computed in 4.8 s and 339159.2 s, respectively. 

Comparing the results from Figs. \ref{fig:sim1_UTD} and \ref{fig:sim2_cutout2}, aberrations appear to be more pronounced along the $z$-dimension or depth. 
This phenomenon is expected, given the analysis in Section \ref{subsec:virtual_array}, where $d_\ell^z$ and the size of the target in the $z$-direction are assumed to be small. 
Hence, for targets of significant size in the $z$-direction, such as the target in Fig. \ref{fig:sim1_UTD_RMA}, the proposed compensation suffers from slight artifacting compared to the BPA. 
However, for many applications, the considerable time savings achieved using our technique is a necessary trade-off compared with the prohibitively slow BPA. 

\subsection{Empirical Irregular Geometry SAR Imaging Results}
\label{subsec:real}
\begin{figure}[h]
\centering
    \begin{subfigure}[b]{0.3\textwidth}
         \centering
         \includegraphics[width=\textwidth]{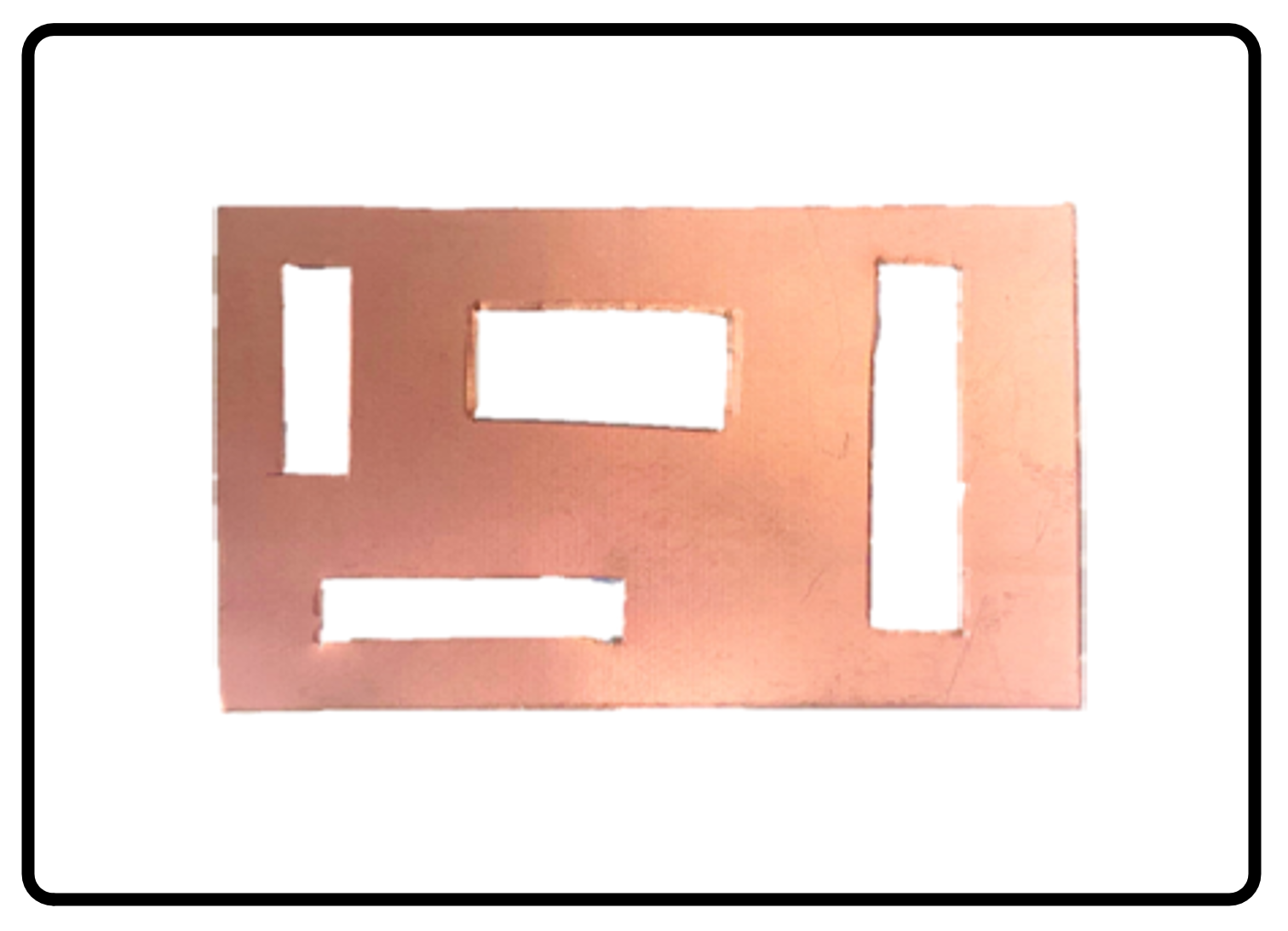} 
         \caption{}
         \label{fig:cutout2}
    \end{subfigure}
    \begin{subfigure}[b]{0.3\textwidth}
         \centering
         \includegraphics[width=\textwidth]{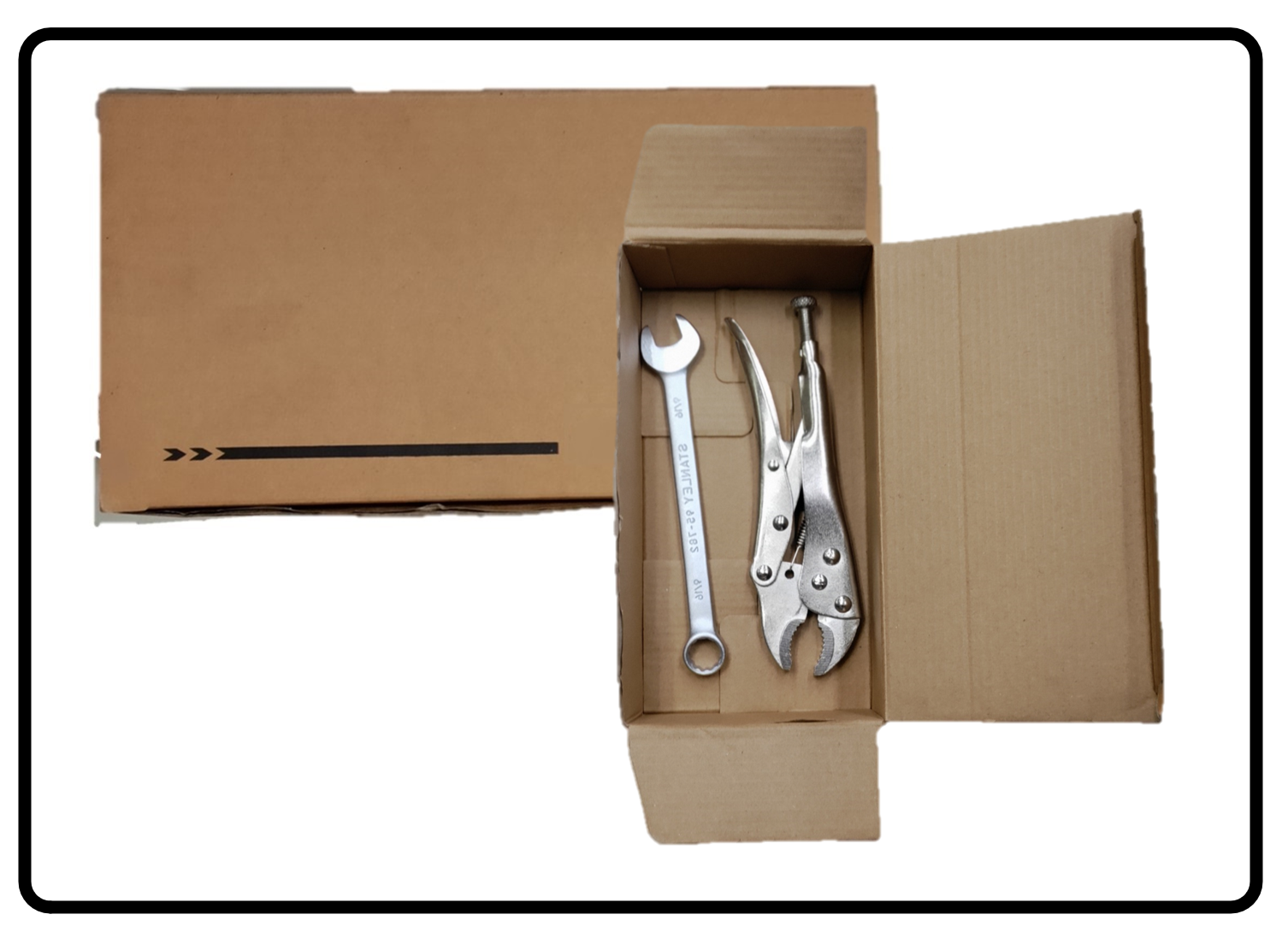} 
         \caption{}
         \label{fig:hiddentools}
    \end{subfigure}
    \vskip\baselineskip
    \begin{subfigure}[b]{0.3\textwidth}
         \centering
         \includegraphics[width=\textwidth]{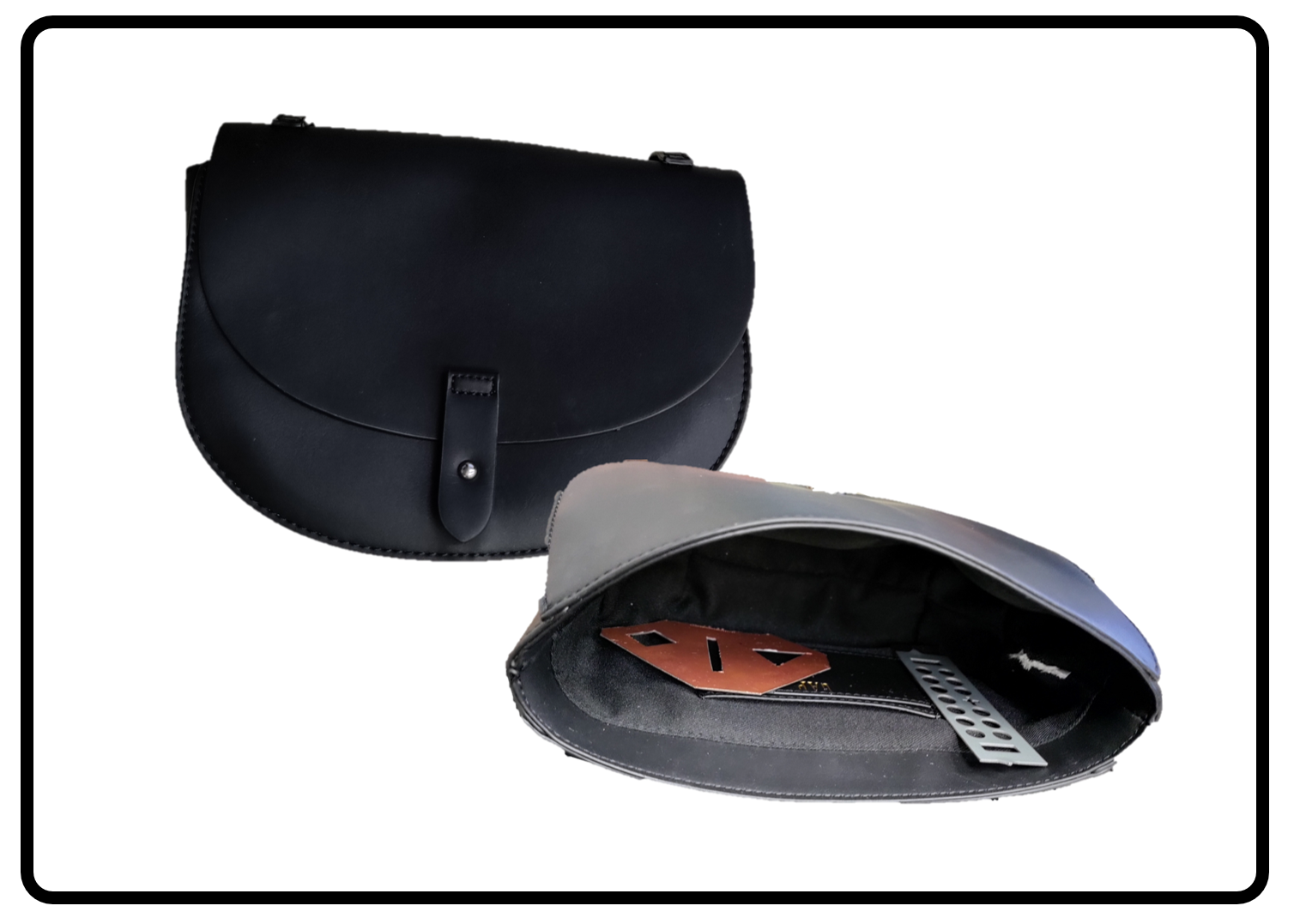} 
         \caption{}
         \label{fig:purse}
    \end{subfigure}
\caption{Various experimental targets: (a) copper clad laminate test target, (b) tools hidden inside box, and (c) purse containing metal cutouts.}
\label{fig:targets}
\end{figure}

The multi-planar multistatic imaging technique and system prototype are validated experimentally by capturing SAR data of various target scenes, as shown in Fig \ref{fig:targets}.
The reconstructed images obtained using each method are compared and discussed.

\begin{figure}[h]
\centering
    \begin{subfigure}[b]{0.35\textwidth}
         \centering
         \includegraphics[width=\textwidth]{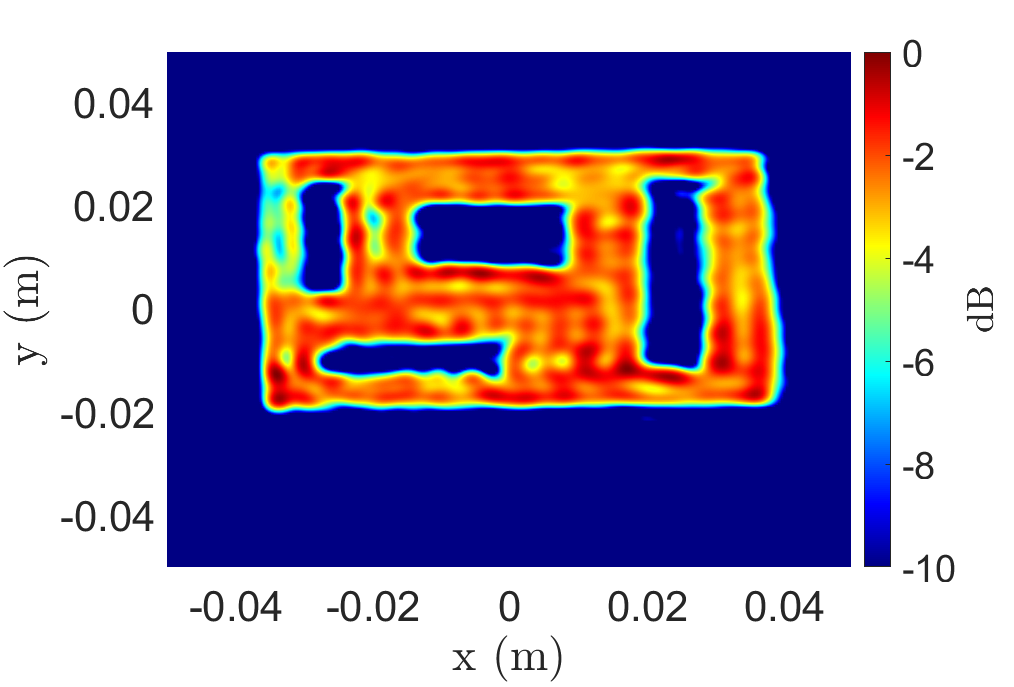} 
         \caption{}
         \label{fig:exp1_cutout2_BPA}
    \end{subfigure}
    \begin{subfigure}[b]{0.35\textwidth}
         \centering
         \includegraphics[width=\textwidth]{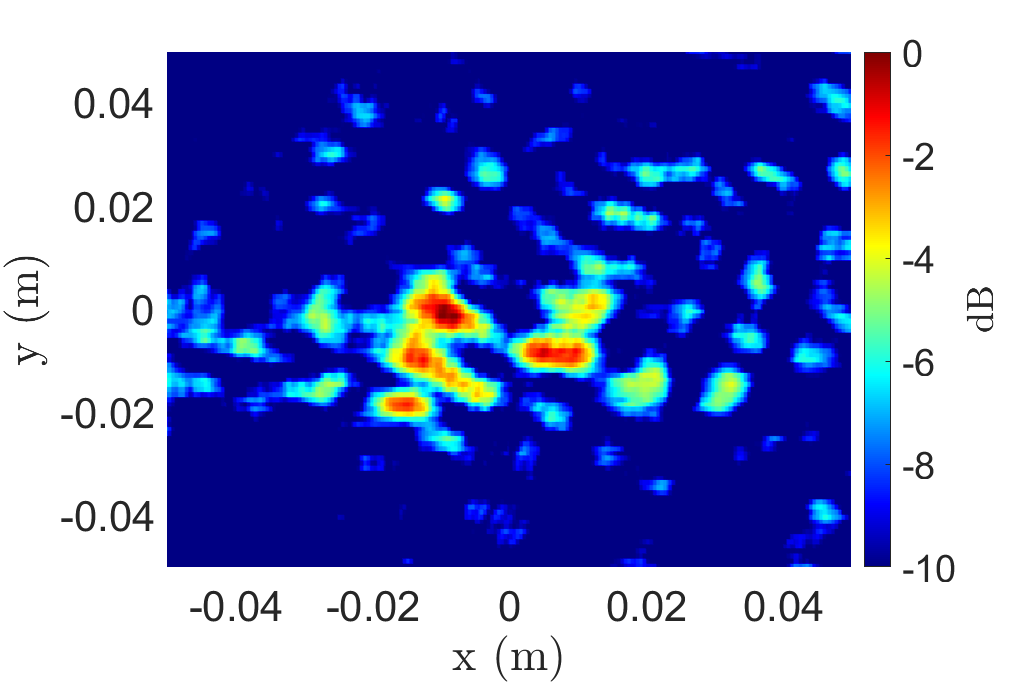} 
         \caption{}
         \label{fig:exp1_cutout2_RMA}
    \end{subfigure}
    \vskip\baselineskip
    \begin{subfigure}[b]{0.35\textwidth}
         \centering
         \includegraphics[width=\textwidth]{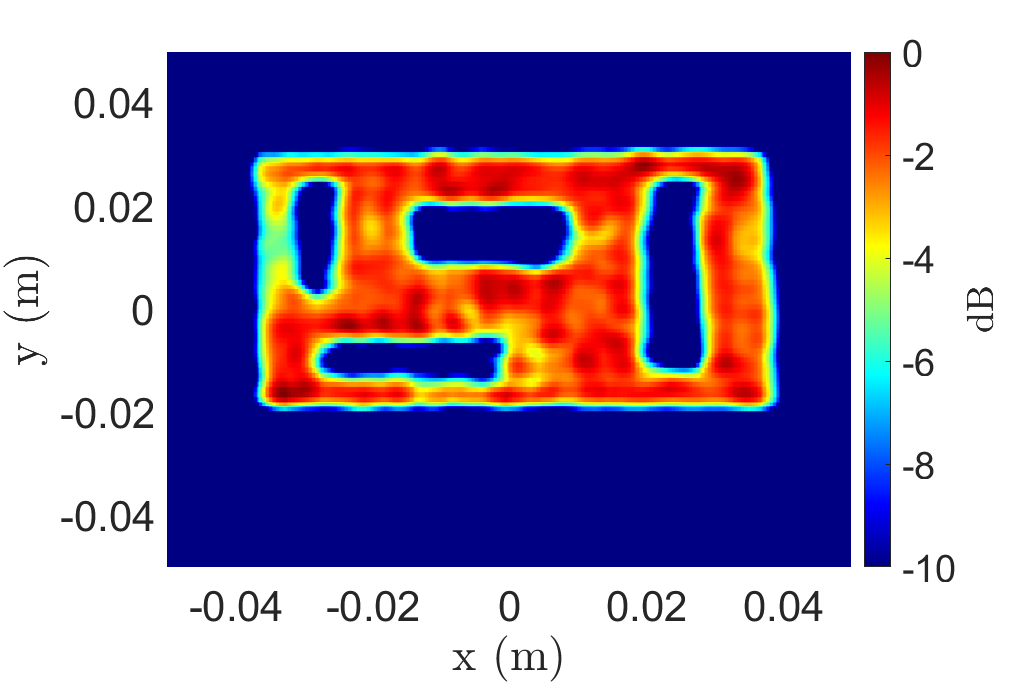} 
         \caption{}
         \label{fig:exp1_cutout2_RMA_FFH}
    \end{subfigure}
\caption{Imaging results for the copper test target using the (a) BPA \mbox{(1324.8 s)}, (b) RMA without multi-planar multistatic compensation \mbox{(1.1 s)}, and (c) proposed algorithm \mbox{(1.1 s)}.}
\label{fig:exp1_cutout2}
\end{figure}

The test target with several horizontal and vertical rectangular cutouts made from copper-clad laminate (Fig. \ref{fig:cutout2}) is illuminated by the $x$-$y$ scanner at the planes \mbox{$z \in [275, 324]$ mm} with a separation of \mbox{1 mm}. 
Hence, data are collected throughout the same region discussed previously such that \mbox{$x' \in [-12.5,12.5]$ cm}, \mbox{$y' \in [-12.5,12.5]$ cm}, and \mbox{$z_\ell \in [-2.5,2.5]$ cm} with 102762 sampling locations. 
After the data were collected, the 50 planar scans were subsampled using a similar random \mbox{2-D} curves as shown in Fig. \ref{fig:sim2_cutout2_scenario}, to emulate the multi-planar irregular sampling scenario.
The imaging results and corresponding computation times for each reconstruction algorithm are shown in Fig. \ref{fig:exp1_cutout2}.
Our proposed multi-planar multistatic imaging technique demonstrates robustness in projecting the irregular scanning geometry to a planar scenario for more efficient image recovery, as the cutout is recovered cleanly without significant artifacting, as shown in Fig. \ref{fig:exp1_cutout2_RMA_FFH}.
In contrast, the image recovered using the BPA requires nearly 30 min to compute, and although the RMA is computed efficiently, the RMA without multi-planar compensation cannot resolve the target scene, as shown in Fig. \ref{fig:exp1_cutout2_RMA}.
Furthermore, when the target location is unknown in the $z$-direction, \mbox{3-D} imaging offers an improved solution with a slightly higher computational cost, using the proposed algorithm.

\begin{figure*}[h]
\centering
    \begin{subfigure}[b]{0.245\textwidth}
         \centering
         \includegraphics[width=\textwidth]{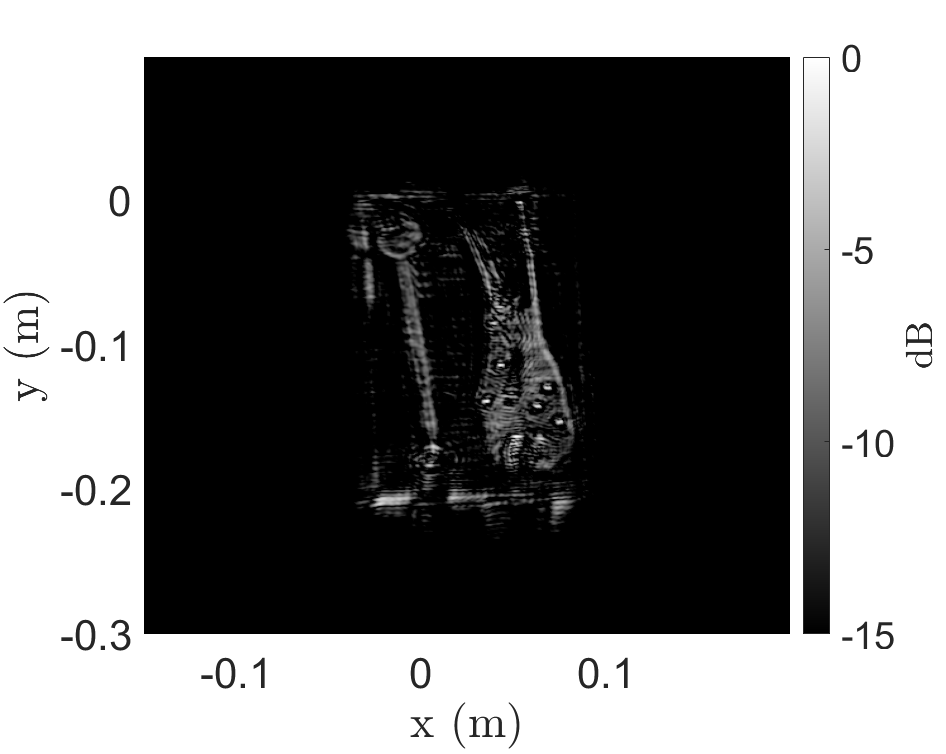} 
         \caption{}
         \label{fig:exp2_hiddentools_BPA_2D}
    \end{subfigure}
    \hfill
    \begin{subfigure}[b]{0.245\textwidth}
         \centering
         \includegraphics[width=\textwidth]{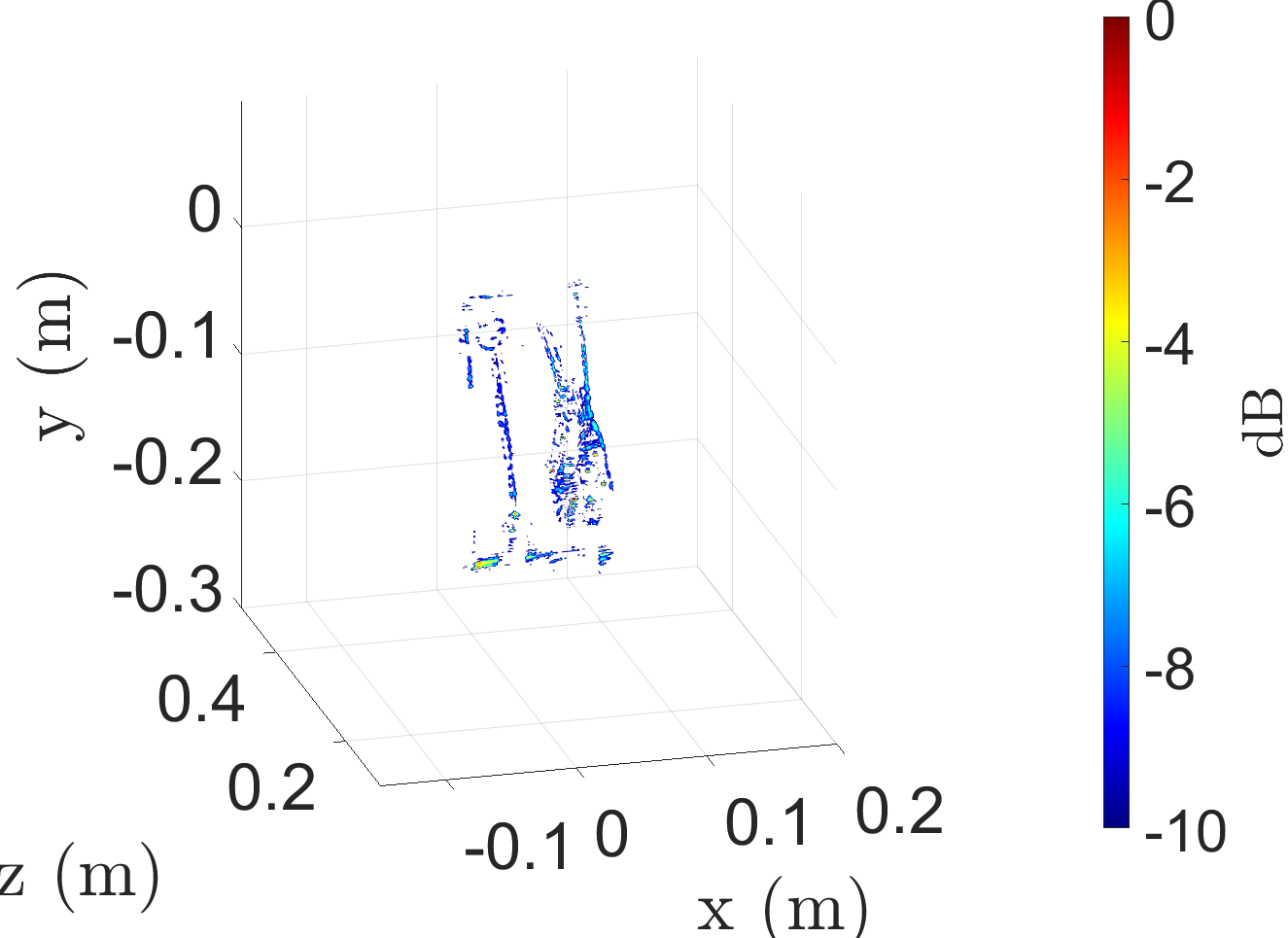} 
         \caption{}
         \label{fig:exp2_hiddentools_BPA_3D}
    \end{subfigure}
    \hfill
    \begin{subfigure}[b]{0.245\textwidth}
         \centering
         \includegraphics[width=\textwidth]{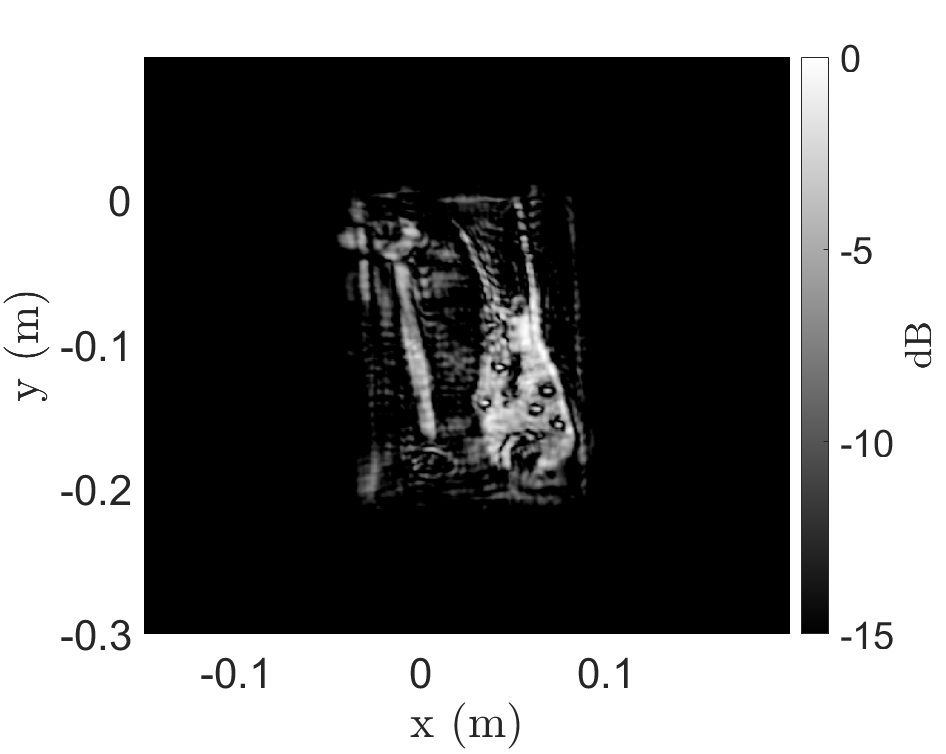} 
         \caption{}
         \label{fig:exp2_hiddentools_RMA_FFH_2D}
    \end{subfigure}
    \hfill
    \begin{subfigure}[b]{0.245\textwidth}
         \centering
         \includegraphics[width=\textwidth]{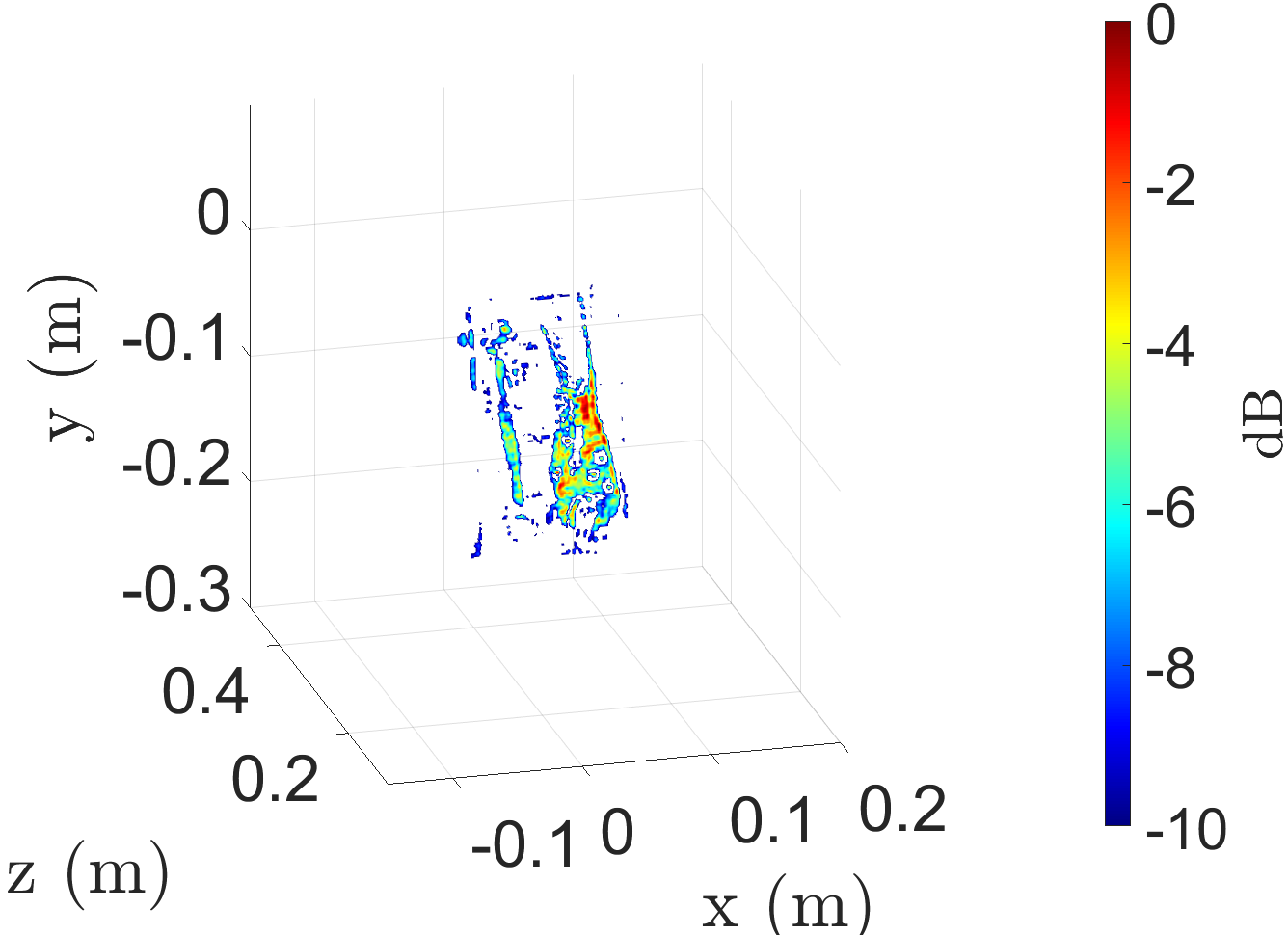} 
         \caption{}
         \label{fig:exp2_hiddentools_RMA_FFH_3D}
    \end{subfigure}
    \vskip\baselineskip
    \begin{subfigure}[b]{0.245\textwidth}
         \centering
         \includegraphics[width=\textwidth]{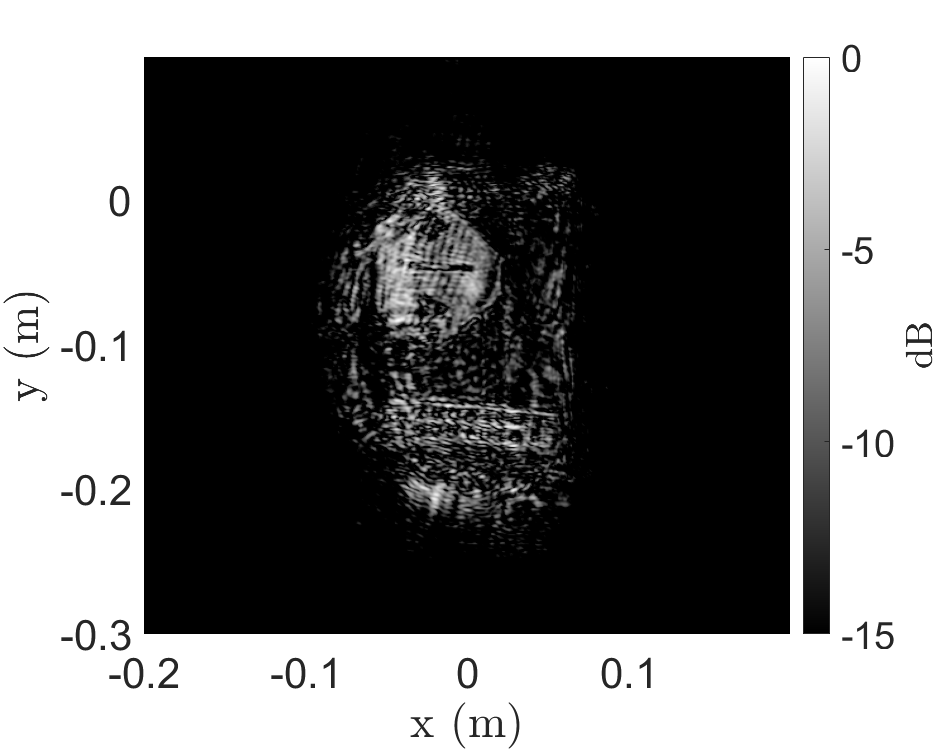} 
         \caption{}
         \label{fig:exp3_purse_BPA_2D}
    \end{subfigure}
    \hfill
    \begin{subfigure}[b]{0.245\textwidth}
         \centering
         \includegraphics[width=\textwidth]{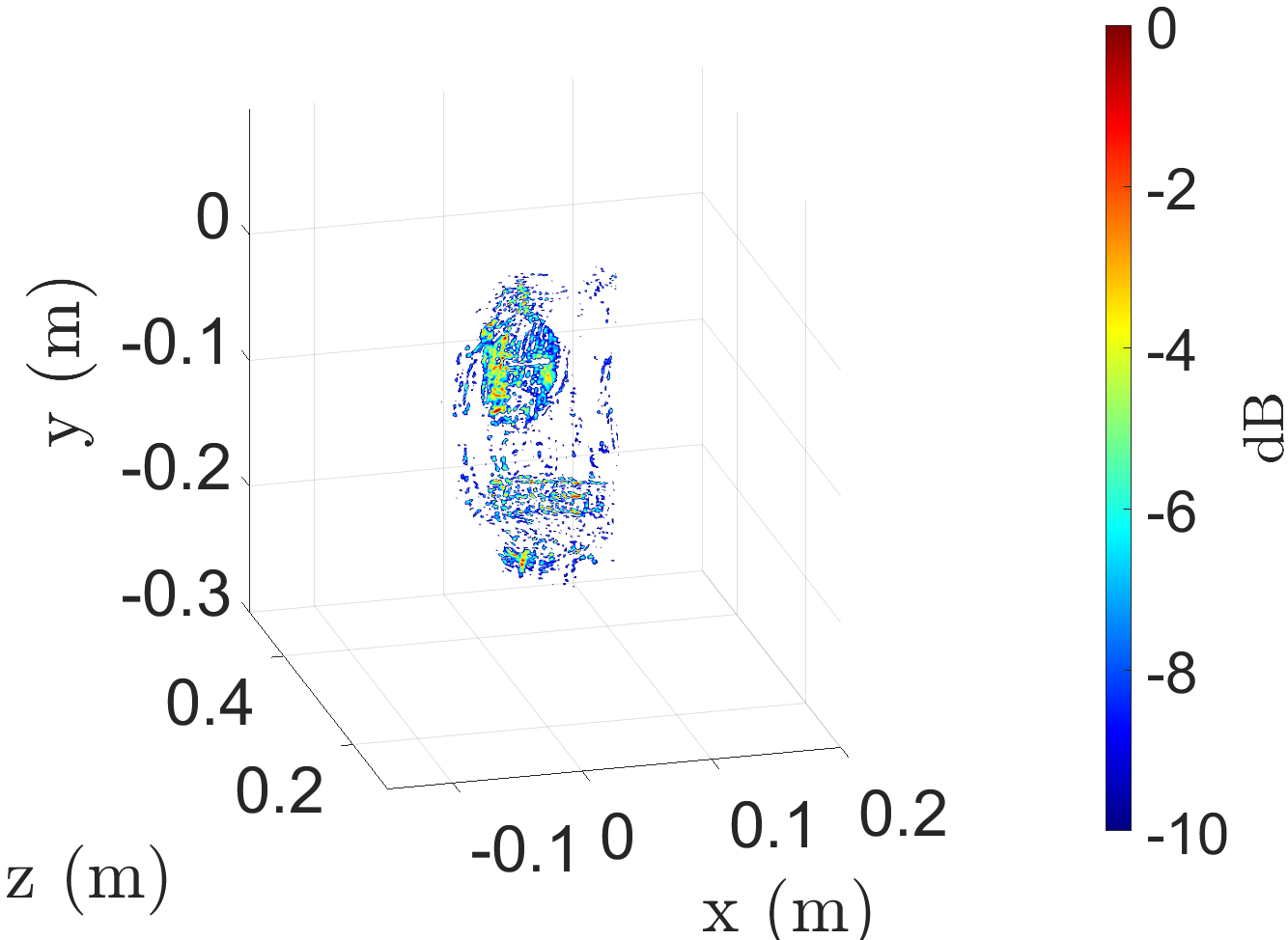} 
         \caption{}
         \label{fig:exp3_purse_BPA_3D}
    \end{subfigure}
    \hfill
    \begin{subfigure}[b]{0.245\textwidth}
         \centering
         \includegraphics[width=\textwidth]{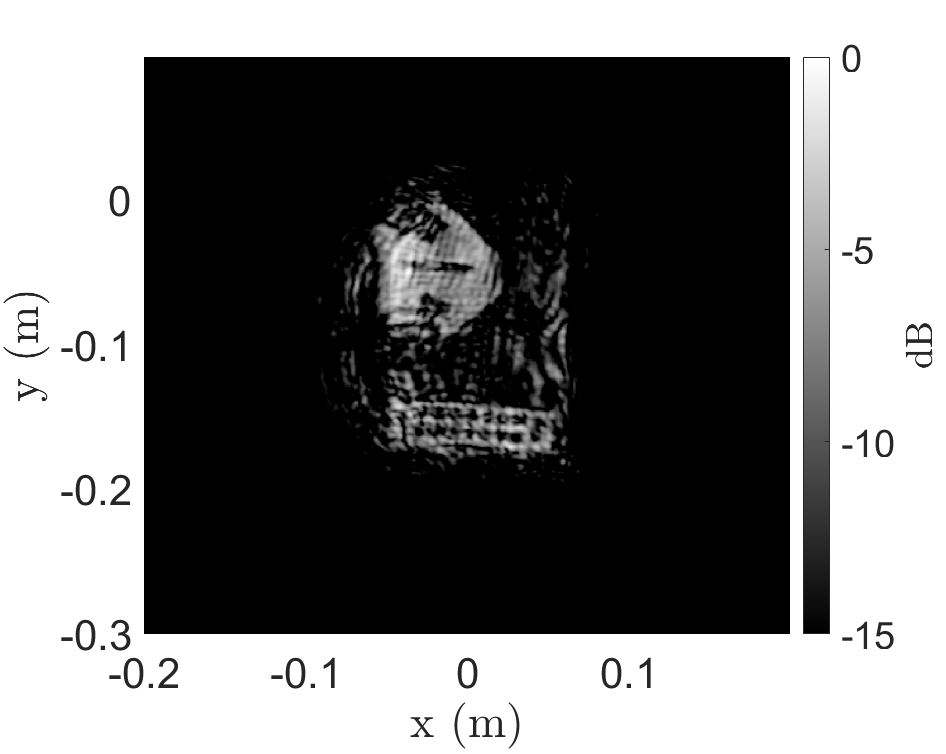} 
         \caption{}
         \label{fig:exp3_purse_RMA_FFH_2D}
    \end{subfigure}
    \hfill
    \begin{subfigure}[b]{0.245\textwidth}
         \centering
         \includegraphics[width=\textwidth]{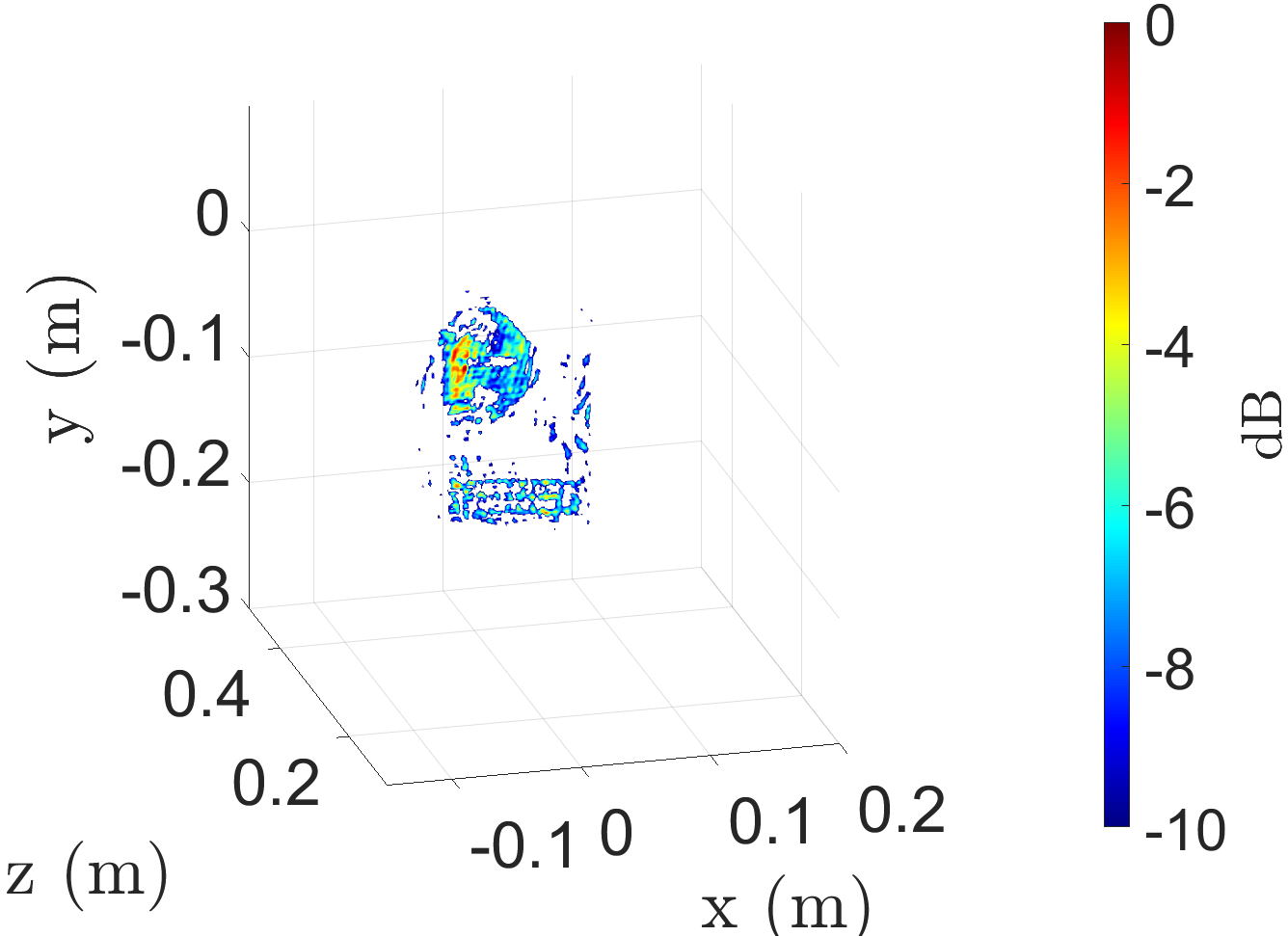} 
         \caption{}
         \label{fig:exp3_purse_RMA_FFH_3D}
    \end{subfigure}
\caption{Imaging results for the hidden tools target, as shown in Fig. \ref{fig:hiddentools}, using the (a) \mbox{2-D} BPA (\mbox{5299.4 s}), (b) \mbox{3-D} BPA (\mbox{1356636.9 s}), (c) \mbox{2-D} proposed algorithm (\mbox{4.3 s}), and (d) \mbox{3-D} proposed algorithm (\mbox{10.7 s}). Imaging results for the concealed items in a purse, as shown in Fig. \ref{fig:purse}, using the (e) \mbox{2-D} BPA (\mbox{5299.4 s}), (f) \mbox{3-D} BPA (\mbox{1356636.9 s}), (g) \mbox{2-D} proposed algorithm (\mbox{4.3 s}), and (h) \mbox{3-D} proposed algorithm (\mbox{10.7 s}).}
\label{fig:exp2_and_exp3}
\end{figure*}

The second target screened by the prototype to demonstrate a hidden target scenario consists of two wrenches (a combination wrench and a vise grip) hidden inside a cardboard box, as shown in Fig. \ref{fig:hiddentools}.
SAR scans of the target are performed with the target at the $z$-planes \mbox{$z \in [275, 324]$ mm} with a separation of \mbox{1 mm}.
To accommodate a larger target size, the aperture is increased to \mbox{$x' \in [-25,25]$ cm}, \mbox{$y' \in [-25,25]$ cm}, and \mbox{$z_\ell \in [-2.5,2.5]$ cm} with 102545 sampling locations.
Similarly, the $x$-$y$-$z$ data are sampled to emulate the multi-planar multi-static scenario using a semi-smooth random curve, as shown in Fig. \ref{fig:sim2_cutout2_scenario}.
The \mbox{2-D} and \mbox{3-D} implementations of the BPA and proposed algorithm are applied to the nonuniform data under the irregular scanning geometry, and the recovered images are shown in Figs. \ref{fig:exp2_hiddentools_BPA_2D} -- \ref{fig:exp2_hiddentools_RMA_FFH_3D}.
Both wrenches are visible in the reconstructed images; however, while the \mbox{2-D} image from the BPA and proposed algorithm provide high-fidelity reconstructions of the hidden tools, the \mbox{2-D} $z$-plane must be carefully selected to obtain such images.
The presence and location of targets are generally unknown for concealed item detection problems. 
Hence, \mbox{3-D} imaging is preferable for such scenarios and is primarily constrained by computational expense.
Our proposed algorithm offers an elegant compromise between the efficiency of the RMA and the image quality of the BPA.
In Fig. \ref{fig:exp2_hiddentools_RMA_FFH_3D}, the \mbox{3-D} image is computed by the proposed algorithm with an image quality comparable to that of the BPA, with a significantly reduced computational cost. 

A third experiment is conducted with several metal cutouts concealed in a purse to emulate a scenario wherein a suspicious personal item is quickly screened with an irregular scanning geometry, such as freehand SAR or drone imaging.
Fig \ref{fig:purse} shows the purse and two hidden items: a triangularly shaped metal plate with different cutout shapes and a rectangular metal plate with circular holes.
The target is scanned by the multi-planar multistatic prototype discussed in Section \ref{sec:prototype}, and the data are employed to emulate an irregular sampling scenario.
Scans are performed with the target at the $z$-planes \mbox{$z \in [275, 324]$ mm} with a separation of \mbox{1 mm} and an aperture is synthesized within \mbox{$x' \in [-25,25]$ cm}, \mbox{$y' \in [-25,25]$ cm}, and \mbox{$z_\ell \in [-2.5,2.5]$ cm} with 102821 sampling locations.
The reconstructed images and corresponding computation times are shown in Figs. \ref{fig:exp3_purse_BPA_2D} -- \ref{fig:exp3_purse_RMA_FFH_3D}.
Both metal cutouts are resolved using our algorithm, with an image quality comparable to that of the BPA. 
Again, assuming that the contents of the purse are generally unknown, computing the \mbox{3-D} image is preferable for concealed item detection.
To efficiently recover a \mbox{3-D} image with irregularly sampled data, existing inversion techniques require excessive computation time and memory capacity, as shown in Figs. \ref{fig:exp3_purse_BPA_2D} and \ref{fig:exp3_purse_BPA_3D}.
However, our proposed algorithm (Figs. \ref{fig:exp3_purse_RMA_FFH_2D} and \ref{fig:exp3_purse_RMA_FFH_3D}) offers an efficient solution that does not compromise image quality.

These experiments demonstrate the advantages of the proposed algorithm and the limitations of the RMA and BPA.
A comparison of the computation times required for each algorithm is presented in Table \ref{tab:computation_times}.
Applying the RMA directly to the multi-planar data, as shown in Figs. \ref{fig:sim1_UTD_RMA}, \ref{fig:sim2_cutout2_RMA}, and \ref{fig:exp1_cutout2_RMA}, yields significant aberrations to the point of failed reconstruction. 
For this reason, the RMA images are not shown in the other examples. 
When the target is known to be \mbox{2-D} and located at a single known $z$-plane, the 2-D BPA implementation can be computed somewhat efficiently in certain instances by employing a graphics processing unit (GPU) and parallelizing the computation \cite{alvarez2019freehand,alvarez2021freehand,alvarez2021system,alvarez2021freehandsystem,alvarez2021towards}.
However, particularly for mobile applications, access to high-capacity GPUs is rare or size-prohibitive, and such acceleration is infeasible.
Moreover, as the BPA is scaled up to three dimensions, the time and space complexities increase exponentially, requiring excessive computational power and memory.
In many emerging applications, efficient \mbox{3-D} image computation on low-power devices is preferable, if not mandatory, as the precise location of the target is generally unknown.
However, efficient algorithms, such as the RMA, require monostatic, planar assumptions that are unachievable by these applications. 
To enable such technologies, our proposed multi-planar multistatic imaging algorithm efficiently compensates for the irregular scanning geometry by carefully handling the phase of each sample.
This enables image reconstruction under dynamic conditions with computational complexity identical to that of the RMA and image quality comparable to that of the BPA.

\begin{table}[h]
    \centering
    \begin{tabular}{c | c | c | c }
         & Metal Cutout & Hidden Tools & Purse \\
         \hline
         \hline
         \mbox{2-D} BPA & 1324.8 & 5299.4 & 5299.4 \\
         \mbox{3-D} BPA & 339159.2 & 1356636.9 & 1356636.9 \\
         \hline
         \mbox{2-D} RMA & 1.1 & 4.3 & 4.3 \\
         \mbox{3-D} RMA & 4.8 & 10.7 & 10.7 \\
         \hline
         \mbox{2-D} Proposed & 1.1 & 4.3 & 4.3 \\
         \mbox{3-D} Proposed & 4.8 & 10.7 & 10.7 \\
         \hline
         \hline
    \end{tabular}
    \caption{Computation time, in seconds, required by the various algorithms for each experiment.}
    \label{tab:computation_times}
\end{table}

\section{Conclusion}
\label{sec:conclusion}
In this article, we presented a novel approach for high-resolution, efficient \mbox{3-D} near-field SAR imaging for irregular scanning geometries. 
We proposed a multi-planar multistatic framework applicable to a diverse set of applications, including freehand imaging, UAV SAR, and automotive imaging. 
A novel algorithm is proposed to efficiently compensate for irregularly sampled multi-planar multistatic data to equivalent planar monostatic mmWave radar data. 
Our technique extends the traditional RMA by presenting an algorithm for efficiently aligning multi-planar multistatic data to a virtual planar monostatic scenario. 
By projecting the data onto a virtual planar monostatic equivalent array, this method extends the RMA to account for both irregular scanning and MIMO-SAR effects, resulting in high-fidelity focusing. 
The proposed algorithm is valid for common radar signaling techniques in 5G, IoT, smartphones, and automotive applications. 
The simulation results demonstrate the robustness of our approach in the presence of significant spatial deviation among the samples along the $z$-direction. 
Furthermore, we empirically validated the proposed algorithm by using a custom prototype to capture multi-planar multistatic data for several concealed and obscured scenarios.
In both simulation and experimental studies, our algorithm achieves efficient image reconstruction matching the focusing quality of existing techniques while reducing computational complexity by a considerable margin. 

\appendix

\section{Multivariate Taylor Series Expansion}
\label{app:taylor_series}
Consider an infinitely differentiable real-valued function and an open neighborhood around $(u,v,w) = (u_0,v_0,w_0)$. Let $\mathbf{x} = [u \ v \ w]^T$ and $\mathbf{x}_0 = [u_0 \ v_0 \ w_0]^T$. Hence, the multivariate Taylor series expansion of $f(\mathbf{x})$ in the neighborhood of $\mathbf{x}_0$ can be written as
\begin{equation}
\label{eq:taylor_series}
    f(\mathbf{x}) = f(\mathbf{x}_0) + (\mathbf{x}-\mathbf{x}_0)^T \nabla \mathbf{f}(\mathbf{x}_0) 
    + \frac{1}{2!}(\mathbf{x}-\mathbf{x}_0)^T \mathbf{H}(\mathbf{x}_0) (\mathbf{x}-\mathbf{x}_0) + \dots ,
\end{equation}
where $\nabla \mathbf{f}$ is the vector of first derivatives
\begin{equation}
    \nabla \mathbf{f}(\mathbf{x}) = 
    \begin{bmatrix}
    f_u(\mathbf{x}) \\
    f_v(\mathbf{x}) \\
    f_w(\mathbf{x})
    \end{bmatrix},
\end{equation}
and $\mathbf{H(x)}$ is the Hessian matrix of the second derivatives as
\begin{equation}
    \mathbf{H(x)} = 
    \begin{bmatrix}
    f_{uu}(\mathbf{x}) & f_{uv}(\mathbf{x}) & f_{uw}(\mathbf{x}) \\
    f_{vu}(\mathbf{x}) & f_{vv}(\mathbf{x}) & f_{vw}(\mathbf{x}) \\
    f_{wu}(\mathbf{x}) & f_{wv}(\mathbf{x}) & f_{vv}(\mathbf{x}) 
    \end{bmatrix}.
\end{equation}

\subsection{Taylor Series Expansion of Round-Trip Distance for Irregular Scanning Geometries}
\label{app:proof}
The round-trip distance between the $\ell$-th Tx/Rx pair, whose transmitter and receiver elements are located at $(x_T,y_T,z_\ell)$ and $(x_R,y_R,z_\ell)$, respectively, and the scatterer located at $(x,y,z)$ is expressed in (\ref{eq:Rl_of_xT_xR_yT_yR}). 
Substituting (\ref{eq:xT_xR_yT_yR_to_virtual}) and (\ref{eq:zl_to_virtual}) into (\ref{eq:Rl_of_xT_xR_yT_yR}), $R_\ell^{RT}$ can be expressed as a function of the distances between the Tx and Rx elements along the $x$- and $y$-directions, $d_\ell^x$ and $d_\ell^y$, respectively, and displacement along the $z$-direction, $d_\ell^z$:
\begin{multline}
\label{eq:Rl}
    R_\ell^{RT}(d_\ell^x,d_\ell^y,d_\ell^z)
    = \left[(x' - \frac{d_\ell^x}{2} - x)^2 + (y' - \frac{d_\ell^y}{2} - y)^2 + (Z_0 + d_\ell^z - z)^2 \right]^{\frac{1}{2}} \\
    + \left[(x' + \frac{d_\ell^x}{2} - x)^2 + (y' + \frac{d_\ell^y}{2} - y)^2 + (Z_0 + d_\ell^z - z)^2 \right]^{\frac{1}{2}}.
\end{multline}
The first derivatives of (\ref{eq:Rl}), evaluated at $d_\ell^x = d_\ell^y = d_\ell^z = 0$, are
\begin{align}
\begin{split}
\label{eq:first_derivatives}
    \frac{\partial R_\ell^{RT}}{\partial d_\ell^x} \biggr\rvert_{(d_\ell^x = d_\ell^y = d_\ell^z = 0)} &= \frac{\partial R_\ell^{RT}}{\partial d_\ell^y} \biggr\rvert_{(d_\ell^x = d_\ell^y = d_\ell^z = 0)} = 0, \\
    \frac{\partial R_\ell^{RT}}{\partial d_\ell^y} \biggr\rvert_{(d_\ell^x = d_\ell^y = d_\ell^z = 0)} &= \frac{2(Z_0 - z)}{R_0},
\end{split}
\end{align}
where $R_0$ is the distance between the virtual monostatic element located at $(x',y',Z_0)$ and the point scatterer at $(x,y,z)$, as expressed in (\ref{eq:R0}).

The second derivatives of (\ref{eq:Rl}), evaluated at the point of interest, can be derived as
\begin{align}
\begin{split}
    \label{eq:second_derivates}
    \frac{\partial^2 R_\ell^{RT}}{\partial (d_\ell^x)^2} \biggr\rvert_{(d_\ell^x = d_\ell^y = d_\ell^z = 0)} &= \frac{1}{2R_0} \left[ 1 - \frac{(x'-x)^2}{R_0^2} \right], \\
    \frac{\partial^2 R_\ell^{RT}}{\partial (d_\ell^y)^2} \biggr\rvert_{(d_\ell^x = d_\ell^y = d_\ell^z = 0)} &= \frac{1}{2R_0} \left[ 1 - \frac{(y'-y)^2}{R_0^2} \right], \\
    \frac{\partial^2 R_\ell^{RT}}{\partial (d_\ell^z)^2} \biggr\rvert_{(d_\ell^x = d_\ell^y = d_\ell^z = 0)} &= \frac{2}{R_0} \left[ 1 - \frac{(Z_0-z)^2}{R_0^2} \right], \\
    \frac{\partial^2 R_\ell^{RT}}{\partial d_\ell^x d_\ell^y} \biggr\rvert_{(d_\ell^x = d_\ell^y = d_\ell^z = 0)} &= -\frac{(x'-x)(y'-y)}{2R_0^3}, \\
    \frac{\partial^2 R_\ell^{RT}}{\partial d_\ell^x d_\ell^z} \biggr\rvert_{(d_\ell^x = d_\ell^y = d_\ell^z = 0)} &= \frac{\partial^2 R_\ell^{RT}}{\partial d_\ell^y d_\ell^z} \biggr\rvert_{(d_\ell^x = d_\ell^y = d_\ell^z = 0)} = 0.
\end{split}
\end{align}

Substituting (\ref{eq:first_derivatives}) and (\ref{eq:second_derivates}) into (\ref{eq:taylor_series}), the quadratic approximation of $R_\ell$ can be expressed as
\begin{multline}
\label{Rl_approximation}
    R_\ell^{RT}\approx 2R_0 + \frac{2(Z_0-z)d_\ell^z}{R_0} + \frac{(d_\ell^x)^2 + (d_\ell^y)^2 + 4(d_\ell^z)^2}{4R_0}
    - \frac{\left[(x'-x)d_\ell^x + (y'-y)d_\ell^y\right]^2 + 4(Z_0-z)^2(d_\ell^z)^2}{4R_0^3}.
\end{multline}

\printbibliography

\end{document}